\documentclass[prd,%
preprint,%
tightenlines,
showpacs,nofootinbib,eqsecnum,amsfonts,amsmath,amssymb]{revtex4}
\usepackage{bm}
\usepackage{graphicx}

\allowdisplaybreaks

\begin{document}

\title{Higher-order spin effects in the dynamics of compact binaries
\\I. Equations of motion}

\author{Guillaume Faye$^{a}$, Luc Blanchet$^{a}$ and Alessandra
Buonanno$^{b,c,a}$} \affiliation{$^a$ ${\mathcal{G}}{\mathbb{R}}
\varepsilon{\mathbb{C}}{\mathcal{O}}$, Institut d'Astrophysique de
Paris, UMR 7095 CNRS Universit\'e Pierre \& Marie Curie,
98$^{\text{bis}}$ boulevard Arago, 75014 Paris, France\\ $^{b}$
Department of Physics, University of Maryland, College Park, MD 20742 \\
$^c$ AstroParticule et Cosmologie (APC), UMR 7164-CNRS, 11, place
Marcellin Berthelot, 75005 Paris, France}

\begin{abstract}
  We derive the equations of motion of spinning compact binaries
  including the spin-orbit (SO) coupling terms one post-Newtonian (PN)
  order beyond the leading-order effect. For black holes maximally
  spinning this corresponds to 2.5PN order. Our result for the equations
  of motion essentially confirms the previous result by Tagoshi, Ohashi
  and Owen. We also compute the spin-orbit effects up to 2.5PN order in
  the conserved (Noetherian) integrals of motion, namely the energy, the
  total angular momentum, the linear momentum and the center-of-mass
  integral. We obtain the spin precession equations at 1PN order beyond
  the leading term, as well. Those results will be used in a future
  paper to derive the time evolution of the binary orbital phase,
  providing more accurate templates for LIGO-Virgo-LISA type
  interferometric detectors.
\end{abstract}


\maketitle

\section{Introduction}\label{secI}

The laser interferometer gravitational-wave (GW) detectors LIGO, Virgo,
GEO 600 and TAMA300 are currently searching for GWs emitted by inspiralling
compact binaries composed of neutron stars and/or black holes. Analyzing
the data using matched filtering technique requires a high-precision
modelling of the inspiral waveform~\cite{3mn, CF94, TNaka94, TSasa94,
DIS98, Bliving, BCV03a, BCV03b, AISS05}. The post-Newtonian (PN)
approximation to general relativity has been applied to build accurate
theoretical templates up to the 3.5PN precision level\,\footnote{As
usual $n$PN refers to terms of order $(v/c)^{2n}$ where $v$ is the
internal velocity and $c$ the speed of light. In this paper we
explicitely display all powers of $c$ and of Newton's constant $G$.} for
\textit{non-spinning} compact bodies~\cite{BFIJ02, ABIQ04, BDEI04}.
Post-Newtonian templates are currently used in analysing the data with
ground-based detectors and in the future they will be used to detect GWs
emitted by supermassive black-hole binaries with the space-based
detector LISA.

Astrophysical observations suggest that black holes can have
non-negligible spins, \textit{e.g}., due to spin up driven by accretion
from a companion during some earlier phase of the binary evolution. For
a few black holes surrounded by matter, observations indicate a
significant intrinsic angular momentum (see, \textit{e.g.},
Refs.~\cite{AK01, Stro01, GD04} for stellar black holes and
Refs.~\cite{FM05, LB03} for supermassive black holes). The spin may even
be close to its maximal value~\cite{TNFIODHIKKMMM95}. Very little is
known however about the black-hole spin magnitudes in binary
systems~\cite{SKKB05}.

To successfully detect GWs emitted by spinning, precessing binaries and to
estimate the binary parameters, spin effects should be included in the
templates. For maximally spinning compact bodies the spin-orbit coupling
(linear in the spins) appears dominantly at the 1.5PN order, while the
spin-spin one (which is quadratic) appears at 2PN order. The spin effect on
the free motion of a test particle was first obtained in the form of a
coupling to curvature by Papapetrou \textit{et al.}~\cite{Papa51, Papa51spin,
  CPapa51spin}. Seminal works by Barker and O'Connell~\cite{BOC75, BOC79}
yielded both the leading order spin-orbit and spin-spin contributions in the
PN equations of motion. More recently, using an effective field theory approach
~\cite{GR06}, leading spin-orbit and spin-spin couplings in the two-body
Hamiltonian were re-derived~\cite{Porto06} and predictions for spin-spin
couplings at 3PN order in the spin potential were obtained~\cite{PR06}.
Based on the works~\cite{BOC75, BOC79}, Kidder, Will and Wiseman~\cite{KWWi93,
  K95} (see also Refs.~\cite{Ger99,MVGer05}) computed the corresponding
coupling terms in the radiation field, enabling thereby the derivation of the
orbital phase evolution, the latter being the crucial quantity that determines
the templates. Currently, only the leading order spin effects, \textit{i.e.},
the spin-orbit and spin-spin couplings have been implemented in the templates
for spinning, precessing black-hole binaries~\cite{ACST94, BCV03b, PBCV04,
  BCPV04, BCPTV05}.

More recently, Tagoshi, Ohashi and Owen~\cite{OTO98, TOO01} started the
computation of the 1PN corrections to the leading spin-orbit coupling.
Those corrections, linear in the spins, appear at 2.5PN order. However,
their work has never been completed: the very important conserved
integrals associated to the equations of motion at 2.5PN order and the
mass quadrupole moment at the 2.5PN order were not computed.

The aim of the present paper together with its
companion~\cite{BBF06spin} is to complete the work of Refs.~\cite{OTO98,
TOO01} and get the orbital phase evolution at 2.5PN order. In this paper
we derive the equations of motion, confirming the main result of
Ref.~\cite{TOO01} (but correcting several important misprints) and
compute the entire set of conserved Noetherian integrals of the motion
associated with the Poincar\'e invariance, notably the energy and the
total angular momentum. In Ref.~\cite{BBF06spin} (henceforth paper~II)
we evaluate the multipole moments and the radiation field so as to
deduce the orbital phase evolution.

The spin of a rotating body is of the order $S^\mathrm{true}\sim
m\,a\,v_\mathrm{spin}$, where $m$ and $a$ denote the mass and typical
size of the body respectively, and where $v_\mathrm{spin}$ represents
the velocity of the body's surface. Here, by ``true'', we mean that the
spin we are referring to is not rescaled [as in Eq.~\eqref{S} below]. In
this paper we shall consider bodies which are both \textit{compact},
$a\sim\frac{G\,m}{c^2}$, and \textit{maximally rotating},
$v_\mathrm{spin}\sim c$. For such objects the magnitude of the spin is
roughly $S^\mathrm{true}\sim\frac{G\,m^2}{c}$. The previous estimate
shows that the spin goes as \textit{one} power of $1/c$, \textit{i.e.},
from the PN point of view, it is formally of order 0.5PN. Again, such a
counting is appropriate for maximally rotating compact objects. It is
then also customary to introduce a dimensionless spin parameter,
generally denoted by $\chi$, defined by
$S^\mathrm{true}=\frac{G\,m^2}{c}\chi$. In our computation the use of
such parameter $\chi$ is not very convenient because it forces us to
introduce some unwanted powers of the mass in front of the spins. On the
other hand, it is useful to keep track of the correct PN order by
counting all the powers of $1/c$. Accordingly we shall ``artificially''
make explicit the factor $1/c$ in front of the spin by posing
$S^\mathrm{true}=S/c$ where $S$ will be considered to be of
``Newtonian'' order. Hence, we shall denote the spin variable by
\begin{equation}\label{S}
S = c\,S^\mathrm{true} = G\,m^2\,\chi\, .
\end{equation}
Such a notation displays explicitly all powers of $1/c$ for maximally
rotating compact objects. Notably, the spin-orbit (SO) effect always
carries a factor $1/c^3$ in front, so that it is regarded as being of
order 1.5PN, while the spin-spin (SS) effect appears at order 2PN in our
terminology, and the 1PN correction to the spin-orbit is 2.5PN order.
This PN counting for spin effects corresponds to the standard practice
when defining the templates of LIGO/Virgo and LISA detectors (see
Refs.~\cite{BCV03a, BCV03b}).

For slowly rotating compact objects ($v_\mathrm{spin}\ll c$) the spin is
formally of higher order, namely
$S^\mathrm{true}\sim\frac{G\,m^2\,v_\mathrm{spin}}{c^2}\sim 1/c^2$,
hence the spin-orbit and spin-spin couplings are pushed at the 2PN and
3PN levels respectively. The 1PN correction to the spin-orbit manifests
itself at the same level as the spin-spin coupling, namely 3PN. Of
course all the computations in this paper and paper~II~\cite{BBF06spin}
are still valid in the case of slow rotation, but in this case the spin
terms are expected to be numerically smaller, and comparable to
higher-order PN contributions.

This paper is organized as follows. In Sec.~\ref{secII} we describe the
stress-energy tensor of spinning point particles and review some
relevant features of the spin formalism before defining our spin
variables. In Sec.~\ref{secIII} we recall some general expressions of
the PN metric and equations of motion, which are valid for arbitrary
extended matter configurations. The PN metric is parametrized by certain
elementary potentials computed in Sec.~\ref{secIV}. Our final results
for the spin-orbit terms in the equations of motion at the 2.5PN order
are presented in Sec.~\ref{secV} in a general frame. They are also
specialized to the center-of-mass frame and reduced to circular orbits.
The precessional equations for the spins including the 1PN relative
correction are derived in Sec.~\ref{secVI}. Finally, in
Sec.~\ref{secVII}, we obtain the spin-orbit contributions to the
conserved integrals associated with our 2.5PN dynamics. The two
Appendices are devoted to some tests of our results.

\section{Stress-energy tensor for spinning point-particles}\label{secII}

Our calculations are based on the standard model of point-particles with
spins~\cite{Papa51, Papa51spin, CPapa51spin, BOC75, BOC79, T57, T59,
Traut58, Dixon, BI80, D82, MST96, TMSS96}. In the Dixon
formulation~\cite{Dixon}, the stress-energy tensor,
\begin{equation}\label{Tmunu}
T^{\mu\nu} = \mathop{T}_\text{M}{}^{\!\mu\nu} +
\mathop{T}_\text{S}{}^{\!\mu\nu}\,,
\end{equation}
is the sum of the ``monopolar'' (M) piece, which is a linear combination
of monopole sources, \textit{i.e.} made of Dirac delta-functions, plus
the ``dipolar'' or spin (S) piece, made of \textit{gradients} of Dirac
delta-functions. The four-dimensional formulation of the monopolar part
reads as
\begin{equation}\label{TmunuM}
\mathop{T}_\text{M}{}^{\!\mu\nu} = c^2 \sum_A
\int_{-\infty}^{+\infty}d\tau_A\,p_A^{(\mu}\,u_A^{\nu)}\,\frac{\delta^{(4)}
(x-y_A)}{\sqrt{-g_A}}\, ,
\end{equation}
where $\delta^{(4)}$ is the four-dimensional Dirac function. The
world-line of particle $A$ (A=1,2), denoted $y_A^\mu$, is parametrized
by the particle's proper time $\tau_A$. The four-velocity is given by
$c\,u_A^\mu=dy_A^\mu/d\tau_A$, and normalized to $g^A_{\mu\nu}u_A^\mu
u_A^\nu=-1$, where $g^A_{\mu\nu}\equiv g_{\mu\nu}(y_A)$ denotes the
metric at the particle's location (the determinant of the metric at
point $A$ being denoted by $g_A$). The four-vector $ p_A^\mu $ is the
particle's linear momentum satisfying
Eqs.~\eqref{eq:momentum}--\eqref{eq:eom} below. The dipolar or spin part
of the stress-energy tensor, which vanishes in the absence of spins,
is\,\footnote{Recall that with our convention the spin variable has the
dimension of a true spin times $c$; the stress-energy tensor has the
dimension of an energy density.}
\begin{equation}\label{TmunuS}
\mathop{T}_\text{S}{}^{\!\mu\nu}=-c\sum_A\nabla_\rho\left[\int_{-\infty}^{
+\infty}d\tau_A\, S_A^{\rho(\mu}\,u_A^{\nu)}\,\frac{\delta^{(4)}
(x-y_A)}{\sqrt{-g_A}}\right]\, ,
\end{equation}
where $\nabla_\rho$ is the covariant derivative associated with the
metric $g_{\mu\nu}$ at the field point $x$, and the anti-symmetric
tensor $ S_A^{\mu\nu} $ represents the spin angular momentum for
particle $A$.

The momentum-like quantity $ p_A^\mu $ is a time-like solution of the
equation
\begin{equation} \label{eq:momentum}
\frac{D S_A^{\mu\nu}}{d\tau_A} \equiv c u_A^\rho \nabla_\rho
S_A^{\mu\nu} = c^2\left(p_A^\mu u_A^\nu - p_A^\nu u_A^\mu\right)\, .
\end{equation}
The equation of motion of the particle with spin, equivalent to the
covariant conservation law of the total stress-energy tensor, namely
$\nabla_\nu T^{\mu\nu}=0$, is given by the Papapetrou
equation~\cite{Papa51, Papa51spin, CPapa51spin}
\begin{equation} \label{eq:eom}
\frac{D p_A^\mu}{d\tau_A} = -\frac{1}{2} S_A^{\lambda \rho} u_A^\nu
R^\mu_{A \,\nu \lambda \rho} \, .
\end{equation}
The Riemann tensor is evaluated at the particle's position $A$,
$R^\mu_{A \,\nu \lambda \rho}\equiv R^\mu_{~\nu \lambda \rho}(y_A)$. The
equation of motion~\eqref{eq:eom} can also be derived directly from the
action principle of Bailey and Israel~\cite{BI80}.

It is well-known that a choice must be made for a supplementary spin
condition (SSC) in order to fix unphysical degrees of freedom associated
with some arbitrariness in the definition of $ S^{\mu\nu} $. This
arbitrariness can be interpreted, in the case of extended bodies, as a
freedom in the choice for the location of the center-of-mass worldline
of the body, with respect to which the angular momentum is defined (see
\textit{e.g.}~\cite{K95} for discussion). In this paper we adopt the
covariant supplementary spin condition
\begin{equation}\label{eq:SSC}
S_A^{\mu\nu}\,p^A_\nu = 0 \, ,
\end{equation}
which allows the natural definition of the spin four-vector $S^A_\mu$ in
such a way that
\begin{equation}\label{spin4}
S_A^{\mu\nu}=-\frac{1}{\sqrt{-g_A}}\,\varepsilon^{\mu\nu\rho\sigma}\,
\frac{p^A_\rho}{m_A c} \,S^A_\sigma \, ,
\end{equation}
where $\varepsilon^{\mu\nu\rho\sigma}$ is the four-dimensional
antisymmetric Levi-Civita symbol such that $\varepsilon^{0123}=1$. For
the spin vector $S^A_\mu$ itself, we choose a four-vector which is
purely spatial in the particle's instantaneous rest frame, where
$u_A^\mu=(1,\mathbf{0})$, hence the components of $S^A_\mu$ are
$(0,\mathbf{S}^A)$ in that frame. Therefore, in any frame,\footnote{The
  alternative choice $ S^A_\mu p_A^\mu=0 $ is equivalent to $ S^A_\mu
  u_A^\mu=0 $ modulo cubic terms in the spins $ \mathcal{O}(S^3) $ (see
  below) which are neglected in the present paper. Such choices are also
  adopted in Refs.~\cite{KWWi93, K95, OTO98, TOO01}.}
\begin{equation}\label{Su}
S^A_\mu u_A^\mu=0 \, .
\end{equation}

As a consequence of the supplementary spin condition~\eqref{eq:SSC}, we
easily verify that $d(S_A^{\mu\nu} S^A_{\mu\nu})/d\tau_A =0$ hence the
spin scalar is conserved along the trajectories: $S_A^{\mu\nu}
S^A_{\mu\nu} =\mathrm{const}$. Furthermore, we can check,
using~\eqref{eq:SSC} and also the Papapetrou law of
motion~\eqref{eq:eom}, that the mass defined by $m_A^2 c^2 = -p_A^\mu
p^A_\mu$ is indeed constant along the trajectories:
$m_A=\mathrm{const}$. Finally, the relation linking the four-momentum
$p_A^\mu$ and the four-velocity $u_A^\mu$ is readily deduced from the
contraction of~\eqref{eq:momentum} with the four-momentum, which results
in
\begin{equation}\label{pu}
p_A^\mu (pu)_A + m_A^2 c^2u_A^\mu = \frac{1}{2c^2} S_A^{\mu \nu}
S_A^{\lambda \rho} u_A^\sigma R^A_{\nu \sigma \lambda \rho}\, ,
\end{equation}
where $(pu)_A\equiv p^A_\nu u_A^\nu$. Contracting further this relation
with the four-velocity one deduces the expression of $(pu)_A$ and
inserting it back into~\eqref{pu} yields the desired relation between
$p_A^\mu$ and $u_A^\mu$.

Let us from now on focus our attention on spin-orbit interactions, which
are \textit{linear} in the spins, and therefore neglect all quadratic
and higher corrections in the spins, say $\mathcal{O}(S^2)$. Drastic
simplifications of the formalism occur in the linear case. Since the
right-hand-side (RHS) of Eq.~\eqref{pu} is quadratic in the spins, we
find that the four-momentum is linked to the four-velocity by the simple
proportionality relation
\begin{equation}\label{pulin}
p_A^\mu = m_A c u_A^\mu + \mathcal{O}(S^2)\, .
\end{equation}
Hence, Eq.~\eqref{eq:SSC} becomes
\begin{equation}\label{eq:SSClin}
S_A^{\mu\nu}\,u^A_\nu = \mathcal{O}(S^3)\, .
\end{equation}
On the other hand, the equation of evolution for the spin, also
sometimes referred to as the \textit{precessional} equation, follows
immediately from the relationship~\eqref{eq:momentum} together with the
law~\eqref{pulin} as $D S_A^{\mu\nu}/d\tau_A = \mathcal{O}(S^2)$, or
equivalently
\begin{equation}\label{parallel}
\frac{D S^A_\mu}{d \tau_A} = \mathcal{O}(S^2)\, .
\end{equation}
This is simply the equation of parallel transport, which means that the
spin vector $S_A^\mu $ remains constant in a freely falling frame, as
could have been expected beforehand. Of course Eq.~\eqref{parallel}
preserves the norm of the spin vector, $S^A_\mu S_A^\mu =
\mathrm{const}$.


When performing PN expansions it is necessary to use three-dimensional
like expressions (instead of four-dimensional) for the stress-energy
tensor. The field point is accordingly denoted by $x=(c\,t,\mathbf{x})$,
and similarly the source points are denoted $y_A=(c\,t,\mathbf{y}_A)$.
The particle trajectories are considered as functions of the coordinate
time $t=x^0/c$, say $\mathbf{y}_A(t)$, and we introduce the ordinary
(coordinate) velocity $v_A^\mu(t)=dy_A^\mu/dt$, also a function of
coordinate time.
Using Eq.~\eqref{pulin} we can write the monopolar part~\eqref{TmunuM}
of the stress-energy tensor as
\begin{equation}
\mathop{T}_\text{M}{}^{\!\mu\nu} = \mathop{T}_\text{NS}{}^{\!\mu\nu} +
\mathcal{O}(S^2)\, ,
\end{equation}
where ${}_\text{NS}T^{\mu\nu}$ is just the standard piece appropriate to
point masses without spins, which reads, in three-dimensional form,
\begin{equation}\label{TmunuM3}
\mathop{T}_\text{NS}{}^{\!\mu\nu} =\sum_A m_A
\frac{v_A^{\mu}v_A^{\nu}}{\sqrt{-g^A_{\rho\sigma}v_A^{\rho}
v_A^{\sigma}/c^2}}\frac{\delta(\mathbf{x}-\mathbf{y}_A)}{\sqrt{-g_A}}\,.
\end{equation}
We have referred to this part of the stress-energy tensor as the
``non-spin'' contribution (NS) in spite of its implicit dependence on
the spins through the metric tensor. Here $\delta\equiv\delta^{(3)}$ is
the three-dimensional Dirac function. Similarly, the spin part of the
stress-energy tensor, Eq.~\eqref{TmunuS}, can be re-written as
\begin{equation}\label{TmunuS3}
\mathop{T}_\text{S}{}^{\!\mu\nu}=-\frac{1}{c}\sum_A\nabla_
\rho\left[S_A^{\rho(\mu}\,v_A^{\nu)}
\frac{\delta(\mathbf{x}-\mathbf{y}_A)}{\sqrt{-g_A}}\right] \, ,
\end{equation}
where the spin tensor $S_A^{\mu\nu}(t)$ is now considered to be a
function of coordinate time, like for the ordinary velocity
$v_A^{\mu}(t)$. The covariant derivative $\nabla_\rho$ acts on
$\mathbf{x}$, which appears in the argument of the delta-function as
shown in~\eqref{TmunuS3}, and on time $t$ through the time-dependence of
the positions $y^\mu_A(t)$, velocities $v^\mu_A(t)$ and spins
$S_A^{\mu\nu}(t)$. 
It is easy to further obtain the more explicit expression
\begin{equation}\label{TmunuS3'}
\sqrt{-g}\,\mathop{T}_\text{S}{}^{\!\mu\nu} = -\frac{1}{c}\sum_A
\biggl\{\partial_\rho\left[S_A^{\rho(\mu}
\,v_A^{\nu)}\,\delta(\mathbf{x}-\mathbf{y}_A)\right] +
S_A^{\rho(\mu}\,\Gamma^{\nu)A}_{\rho\sigma}\,v_A^\sigma
\,\delta(\mathbf{x}-\mathbf{y}_A)\biggr\} \, ,
\end{equation}
where $\Gamma^{\nu A}_{\rho\sigma}\equiv\Gamma^{\nu}_{\rho\sigma}(y_A)$
denotes the Christoffel symbol evaluated at the source point $A$, and
where one should notice that the square-root of the determinant
$\sqrt{-g}$ in the left-hand-side (LHS) is to be evaluated at the
\textit{field} point $(t,\mathbf{x})$, contrarily to the factor
$1/\sqrt{-g_A}$ in the RHS of Eq.~\eqref{TmunuS3} which is to be
computed at the \textit{source} point $y_A=(c t,\mathbf{y}_A)$. The
explicit form~\eqref{TmunuS3'} of the spin stress-energy tensor is used
in all our practical calculations.

In terms of three-dimensional variables the spin tensor reads [after
taking into account the spin condition~\eqref{Su}, namely
$S^A_0=-S^A_iv_A^i/c$]
\begin{subequations}\label{SA3}
\begin{align}
S_A^{0i} &=
-\frac{1}{\sqrt{-g_A}}\,\varepsilon^{ijk}\,u^A_j\,S^A_k
\, , \\ S_A^{ij} &=
-\frac{1}{\sqrt{-g_A}}\,\varepsilon^{ijk}\left[u^A_0\,S^A_k
  +u^A_k \frac{v_A^l}{c}S^A_l\right] \, ,
\end{align}
\end{subequations}
where $\varepsilon^{ijk}$ is the ordinary Levi-Civita symbol such that
$\varepsilon^{123}=1$. Here, we have
\begin{subequations}\label{uA3}
\begin{align}
u^A_0&= u_A^0\left[g_{00}^A+g_{0i}^A\,\frac{v_A^i}{c}\right]\,,\\
u^A_j&=u_A^0 \left[g^A_{j0} + g^A_{jk} \frac{v_A^k}{c}\right]\,,\\
\text{with}~~u_A^0&=
\frac{1}{\sqrt{-g^A_{\rho\sigma}\,v_A^\rho\,v_A^\sigma/c^2}} \, .
\end{align}
\end{subequations}
In principle we could adopt as the basic spin variable the covariant
vector (or covector) $S^A_i$. However, we shall instead use
systematically the \textit{contravariant} components of the vector
$S_A^i$, which are obtained by raising the index on $S^A_k$ by means of
the spatial metric $\gamma_A^{ik}$, which denotes the inverse of the
covariant spatial metric evaluated at point $A$, $\gamma^A_{kj}\equiv
g^A_{kj}$ (\textit{i.e.} such that
$\gamma_A^{ik}\gamma^A_{kj}=\delta^i_j$). Hence we \textit{define} (and
systematically use in all our computations)
\begin{equation}\label{SAcov}
S_A^i\equiv\gamma_A^{ik}S^A_k\quad\Longleftrightarrow\quad
S^A_i\equiv\gamma^A_{ij}S_A^j\, .
\end{equation}
Beware of the fact that the latter definition of the contravariant spin
variable $S_A^i$ differs from the possible alternative choice
$g_A^{i\nu}S^A_\nu$. The spin vector $S_A^i$ as defined by~\eqref{SAcov}
agrees with the choice already made in Refs.~\cite{OTO98, TOO01}.

\section{Post-Newtonian metric and equations of motion}\label{secIII}

The starting point is the general formulation, \textit{i.e.} valid for
any matter stress-energy tensor $T^{\mu\nu}$ with spatially compact
support, of the PN metric and equations of motion at 2.5PN
order, as worked out in Ref.~\cite{BFP98}. In harmonic (or De
Donder) coordinates\,\footnote{Thus
$\partial_\nu\left(\sqrt{-g}\,g^{\mu\nu}\right)=0$, where $g^{\mu\nu}$
is the inverse of the usual covariant metric $g_{\mu\nu}$, and
$g=\mathrm{det}(g_{\rho\sigma})$.} the 2.5PN metric is expressed in
terms of certain ``elementary'' potentials as
\begin{subequations}\label{eq:metric2.5PN}
\begin{align}
g_{00} &= -1 + \frac{2}{c^2}V - \frac{2}{c^4}V^2 + \frac{8}{c^6}
\left[\hat{X} + V_iV_i + \frac{V^3}{6}\right]
+\mathcal{O}\left(\frac{1}{c^8}\right) \, , \\ g_{0i} &= -\frac{4}{c^3}
V_i - \frac{8}{c^5} \hat{R}_i + \mathcal{O}\left(\frac{1}{c^7}\right) \,
,\\ g_{ij} &= \delta_{ij} \left( 1 + \frac{2}{c^2} V + \frac{2}{c^4} V^2
\right) + \frac{4}{c^4} \hat{W}_{ij} +
\mathcal{O}\left(\frac{1}{c^6}\right) \, .
\end{align}
\end{subequations}
These potentials, $V$, $V_i$, $\cdots$, are defined by some retarded
integrals of appropriate PN iterated sources. To define them it is
convenient to introduce the matter source densities
\begin{subequations}\label{sigmamunu}
\begin{align}
 \sigma &= \frac{T^{00}+T^{kk}}{c^2} \, , \\ \sigma_i &=
 \frac{T^{0i}}{c} \, , \\ \sigma_{ij} &= T^{ij}
\end{align}
\end{subequations}
(with $T^{kk}\equiv\delta_{ij}T^{ij}$). Then, with
$\Box^{-1}_\mathrm{R}$ denoting the usual flat space-time retarded
operator, we have for the Newtonian like potential $V$,
\begin{subequations}\label{potentials}
\begin{equation}
V = \Box^{-1}_\mathrm{R}\bigl\{-4\pi G \sigma \bigr\} \equiv G
\int\frac{d^3\mathbf{x}'}{\vert\mathbf{x}-\mathbf{x}'\vert}\,
\sigma\left(\mathbf{x}', t-\vert\mathbf{x}-\mathbf{x}'\vert/c\right) \,
.\label{V}
\end{equation}
The higher-order PN potentials read
\begin{align}
V_i &= \Box^{-1}_\mathrm{R}\bigl\{-4\pi G \sigma_i\bigr\} \, ,\\
\hat{W}_{ij} &= \Box^{-1}_\mathrm{R}\bigl\{-4 \pi G (\sigma_{ij} -
\delta_{ij} \sigma_{kk}) - \partial_i V \partial_j V\bigr\} \, ,
\label{Wij}\\ \hat{R}_i &= \Box^{-1}_\mathrm{R}\Bigl\{ - 4\pi G
(V\sigma_i - V_i \sigma) - 2 \partial_k V \partial_i V_k - \frac{3}{2}
\partial_t V \partial_i V \Bigr\} \, ,\\ \hat{X} &=
\Box^{-1}_\mathrm{R}\Bigl\{ -4\pi G V \sigma_{ii} + 2 V_i \partial_t
\partial_i V +V \partial_t^2 V \nonumber \\ &\quad\quad~+\frac{3}{2}
(\partial_t V)^2 - 2 \partial_i V_j \partial_j V_i + \hat{W}_{ij}
\partial^2_{ij} V \Bigr\} \, .
\end{align}
\end{subequations}
All these potentials are subject, up to the required PN order, to the
differential identities
\begin{subequations}\label{iddiff}
\begin{align}
&\partial_t\left\{V+\frac{1}{c^2}\left[\frac{1}{2}\hat{W}_{ii}
+2V^2\right]\right\}
+\partial_i\left\{V_i+\frac{2}{c^2} \left[\hat{R}_i+VV_i\right]\right\}=
\mathcal{O}\left(\frac{1}{c^4}\right) \, , \\ &\partial_t V_i
+\partial_j\left\{\hat{W}_{ij}-\frac{1}{2}\delta_{ij} \hat{W}_{kk}
\right\}=\mathcal{O}\left(\frac{1}{c^2}\right) \, ,
\end{align}
\end{subequations}
which are consequences of the harmonic coordinate conditions; see
Ref.~\cite{BFP98}.

In this paper we shall specialize the latter PN metric to systems of
particles with spin. In this case, as we have reviewed in
Sec.~\ref{secII}, the stress tensor is the sum of the non-spin piece
given by \eqref{TmunuM3} and of the spin part~\eqref{TmunuS3}, thus
$T^{\mu\nu} = {}_\text{NS}T^{\mu\nu}+{}_\text{S}T^{\mu\nu}$. Henceforth
we often do not indicate the neglected $\mathcal{O}(S^2)$ terms. Hence,
the source densities~\eqref{sigmamunu} will be of the form
$\sigma_{\mu\nu}={}_\text{NS}\sigma_{\mu\nu}+{}_\text{S}\sigma_{\mu\nu}$,
and all the potentials will thus admit similar decompositions, say
\begin{subequations}\label{Vdecomp}
\begin{align}
V&=\mathop{V}_\text{NS} +\mathop{V}_\text{S}\,,~\cdots \, ,\\
\hat{W}_{ij}&=\mathop{\hat{W}}_\text{NS}{}_{\!ij}
+\mathop{\hat{W}}_\text{S}{}_{\!ij}\,,~\cdots \, .
\end{align}
\end{subequations}
%
The equations of motion of spinning particles are obtained from the
covariant conservation of the total stress-energy tensor,
\begin{equation}\label{conslaw}
0 = \nabla_\nu T^{\mu\nu} = \nabla_\nu
\mathop{T}_\text{NS}{}^{\!\mu\nu}+\nabla_\nu\mathop{T}_\text{S}{}^{\!\mu\nu}
+ \mathcal{O}(S^2) \, .
\end{equation}
To get the acceleration of the $A$-th particle, we insert into the
conservation law~\eqref{conslaw} the expressions~\eqref{TmunuM3} and
\eqref{TmunuS3} of the stress tensor, integrate over a small volume
surrounding the particle $A$ (excluding the other particles $B$), and
use the properties of the Dirac delta-function. More precisely, in order
to handle the delta-function, we systematically apply the rules
appropriate to Hadamard's partie finie regularization and given by
Eq.~\eqref{hadamard} below. As a result we obtain the equations of
motion of the particle $A$ and find useful to write them in the form
\begin{equation}\label{dPAdt}
\frac{d P^A_\mu}{d t} = F^A_\mu \, ,
\end{equation}
where both the ``linear momentum density'' $P^A_\mu$ and ``force
density'' $F^A_\mu$ (per unit mass) involve a non-spin piece (NS) and
the spin part (S),
\begin{subequations}\label{PAFA}
\begin{align}
P^A_\mu &=
\mathop{P}_\text{NS}{}^{\!A}_{\!\mu}+\mathop{P}_\text{S}{}^{\!A}_{\!\mu}
\\ \text{and}\quad F^A_\mu &=
\mathop{F}_\text{NS}{}^{\!A}_{\!\mu}+\mathop{F}_\text{S}{}^{\!A}_{\!\mu}
\, .
\end{align}
\end{subequations}
The non-spin parts correspond to the geodesic equations and read
\begin{align}
\mathop{P}_\text{NS}{}^{\!A}_{\!\mu} &= \frac{v_A^\nu
\,g^A_{\mu\nu}}{\sqrt{-g^A_{\rho\sigma}v_A^\rho v_A^\sigma/c^2}}\,, \\
\mathop{F}_\text{NS}{}^{\!A}_{\!\mu} &= \frac{1}{2}\frac{v_A^\nu
\,v_A^\lambda \,(\partial_\mu
g_{\nu\lambda})_A}{\sqrt{-g^A_{\rho\sigma}v_A^\rho v_A^\sigma/c^2}}\,.
\end{align}
Their complete expressions in terms of the elementary potentials
\eqref{potentials} were given in Ref.~\cite{BFP98}. We shall need them
for a spatial index ($\mu=i$) and for completeness we report here the
result (see Eqs.~(8.3) in~\cite{BFP98})
\begin{subequations}\label{exprPAFA}
\begin{align}
\mathop{P}_\text{NS}{}^{\!A}_{\! i} &= v_A^i +
\frac{1}{c^2}\left[-4V_i+3Vv^i+\frac{1}{2}v^2v^i\right]_A \nonumber\\
&+\frac{1}{c^4}\left[-8\hat{R}_i+\frac{9}{2}V^2v^i+4\hat{W}_{ij}v^j-4VV_i
\right.\nonumber\\
&+\left.\frac{7}{2}Vv^2v^i-2v^2V_i-4v^iv^jV_j+\frac{3}{8}v^iv^4\right]_A
+\mathcal{O}\left(\frac{1}{c^6}\right) \, ,\\ \nonumber \\[-3mm]
\mathop{F}_\text{NS}{}^{\!A}_{\! i} &= (\partial_i V)_A +
\frac{1}{c^2}\left[- V \partial_i V + \frac{3}{2} v^2 \partial_i V-
4v^j\partial_i V_j \right]_A \nonumber\\
&+\frac{1}{c^4}\left[4\partial_i\hat{X}+ 8V_j \partial_i V_j - 8
v^j\partial_i \hat{R}_j +\frac{9}{2}v^2V\partial_i V \right.\nonumber\\
&+ \left. 2 v^j v^k \partial_i \hat{W}_{jk} - 2 v^2 v^j\partial_i V_j +
\frac{7}{8} v^4 \partial_i V + \frac{1}{2}V^2 \partial_i V
\right.\nonumber\\ &- \left. 4 v^j V_j\partial_i V - 4v^j V\partial_i
V_j \right]_A + \mathcal{O}\left(\frac{1}{c^6}\right) \, .
\end{align}
\end{subequations}
These expressions are still valid in the present situation, but we have
to remember that the elementary potentials therein do involve
contributions from the spins, \textit{e.g.}
$V={}_\text{NS}V+{}_\text{S}V$. Therefore it is crucial to compute the
spin parts of the potentials and to insert them into the non-spin
(geodesic-like) contributions to the equations of motion,
Eqs.~\eqref{exprPAFA}.

Now the purely spin parts, ${}_\text{S}P^A_\mu$ and
${}_\text{S}F^A_\mu$, will produce a deviation from the geodesic motion
which is induced by the effect of spins. We have found that they admit
the following expressions,
\begin{subequations}\label{PAFAspin}
\begin{align}
m_A c \,\mathop{P}_\text{S}{}^{\!A}_{\!\mu} &=
-\frac{1}{2c}\frac{d}{dt}\left(g^A_{\mu\nu}\,S_A^{0\nu}\right)
+\frac{1}{2}\left(\partial_\rho g_{\mu\nu}\right)_A
\,S_A^{\rho\nu}\nonumber\\& -\frac{1}{2}g^A_{\rho\nu}\,\Gamma^{\nu
  A}_{\mu \sigma}\,S_A^{\rho
  0}\,\frac{v_A^\sigma}{c}-\frac{1}{2}g^A_{\mu\nu}\,\Gamma^{0
  A}_{\rho\sigma}\,S_A^{\rho\nu}\,\frac{v_A^\sigma}{c} \, , \\ m_A
c\,\mathop{F}_\text{S}{}^{\!A}_{\!\mu} &=
\frac{1}{2}\left(\partial_{\mu\rho} g_{\nu\sigma}\right)_A
\,S_A^{\rho\nu}\,v_A^\sigma \nonumber\\& -\frac{1}{2}\left(\partial_\mu
g_{\nu\lambda}\right)_A\,\Gamma^{\nu
  A}_{\rho\sigma}\,S_A^{\rho\lambda}\,v_A^\sigma \, .
\end{align}
\end{subequations}
To compute them is relatively straightforward because all the metric
coefficients and Christoffel symbols therein take their standard
non-spin expressions (since we are looking for an effect linear in the
spins), and these have already been computed in Ref.~\cite{BFP98}.

As a check of our calculations we have also used an alternative
formulation of the equations of motion, which is directly obtained from
the Papapetrou equations of motion~\eqref{eq:eom} and reads, at
linearized order in the spins,
\begin{equation} \label{eq:eomlin}
m_A c \frac{D u_A^\mu}{d\tau_A} = -\frac{1}{2} S_A^{\lambda \rho}
u_A^\nu R^\mu_{A \,\nu \lambda \rho} + \mathcal{O}(S^2)\, .
\end{equation}
We lower the free index $ \mu $ so as to use the convenient relation $D
u^A_\mu/d \tau_A = du_\mu^A/d\tau_A - \frac{1}{2}u_A^\nu u_A^\lambda
(\partial_\mu g_{\nu \lambda})_A $. The resulting equation takes the same
form as Eq.~\eqref{dPAdt},
\begin{equation}\label{dPAdtnew}
\frac{d \mathcal{P}^A_\mu}{d t} = \mathcal{F}^A_\mu \, ,
\end{equation}
but with some distincts linear momentum and force densities
$\mathcal{P}^A_\mu$ and $\mathcal{F}^A_\mu$. It is clear that the
non-spin parts, corresponding to geodesic motion, can be taken to be
exactly the same as in our previous formulation, namely
Eqs.~\eqref{exprPAFA}. However the spin parts are different; they are
given in terms of the Riemann tensor $R^A_{\mu \lambda \sigma \tau}
\equiv R_{\mu \lambda \sigma \tau}(y_A)$ as follows,
\begin{subequations}\label{PAFApapa}
\begin{align} \label{PApapa}
m_A c\,\mathop{\mathcal{P}}_\text{S}{}^{\!A}_{\!\mu} &= 0 \, , \\
\label{FApapa} 
m_A c\,\mathop{\mathcal{F}}_\text{S}{}^{\!A}_{\!\mu} &=
\frac{R^A_{\mu \lambda \sigma \tau} \,\varepsilon^{\nu \rho \sigma
\tau}}{2 \sqrt{g^A \,g^A_{\pi\epsilon} \,v_A^\pi \,v_A^\epsilon}}
\,v_A^\lambda \,g_{\nu \omega}^A \,v_A^\omega \,S^A_\tau \, ,
\end{align}
\end{subequations}
where $S^A_\tau$ is the covariant spin covector appearing
in~\eqref{SAcov}. The difference with Eqs.~(5.1--3) in Tagoshi
\textit{et al.}~\cite{TOO01} is due to the fact that these authors work
on the contravariant version of the Papapetrou equation. The advantage
of the formulation~\eqref{PAFApapa} over the previous
one~\eqref{PAFAspin} is of course that it is manifestly covariant. This
advantage is however relatively minor in practical PN calculations,
since the manifest covariance of the equations is anyway broken from the
start. It remains that the two formulations are very useful, and their
joint use provides a very good check of the calculations.

The quantity~\eqref{FApapa} can be computed from the 2.5PN metric,
by inserting it into the curvature tensor $R^A_{\mu \lambda \sigma
\tau}$, but we may also express them directly by means of the elementary
potentials~\eqref{potentials}. Let us give here the complete result at
the required PN order,
\begin{align}\label{PAFAresult}
m_A c\,\mathop{\mathcal{F}}_\text{S}{}^{\!A}_{\!i} &= \frac{1}{c^3}
\left\{\varepsilon_{ijk} \Big(\partial_j \partial_t V + v^l
\partial_{jl} V\Big) S^k + 2 \varepsilon_{jkl} \partial_{il} \Big(V v^j
- V_j\Big) S^k \right\}_A \nonumber\\ &+ \frac{1}{c^5}
\Bigg\{\varepsilon_{ijk} \bigg[ \Big(\partial_j \partial_t V + v^l
\partial_{jl} V\Big) \Big(S^k V + \frac{1}{2} v^2 S^k - (Sv) v^k \Big) +
\Big(\partial^2_t V + v^l \partial_l \partial_t V\Big) v^j S^k
\nonumber\\ & \qquad \qquad + 2 \partial_l V \Big(\partial_l V_k S^j +
S^k \partial_j V_l + v^j S^k \partial_l V\Big) + \partial_j V
\Big(\partial_t V - v^l \partial_l V\Big) S^k \bigg] \nonumber\\ &
\qquad + \varepsilon_{jkl} \bigg[2 \Big(2 \partial_j V \partial_l V_i -
2 V_j \partial_{il} V + 2 \partial_i V \partial_l V_j + V \partial_{il}
V_j \nonumber\\ & \qquad \qquad - 2 v^j \partial_l V \partial_i V + v^j
V \partial_{il} V + v^j v^m \partial_{lm} V_i - v^j v^m \partial_{il}
V_m \nonumber\\ & \qquad \qquad + v^j \partial_l \partial_t V_i - v^l
\partial_i \partial_t V_j + v^m \partial_{il} \hat{W}_{jm} - v^m
\partial_{lm} \hat{W}_{ij} - \partial_l \partial_t \hat{W}_{ij} - 2
\partial_{il} \hat{R}_j\Big) S^k \nonumber\\ & \qquad \qquad + v^2
\Big(- \partial_{il} V_j + v^j \partial_{il} V\Big) S^k + 2 (Sv) v^k
\partial_{il} V_j \bigg]\Bigg\}_A \, .
\end{align}

\section{Computation of the spin parts of elementary potentials}\label{secIV}

We shall compute all the spin parts of the elementary potentials listed
in Eqs.~\eqref{potentials}, which are needed for insertion into the
``non-spin'' parts of the momentum and force densities as defined by
Eq.~\eqref{exprPAFA}. Here we do not compute the non-spin parts of the
potentials since they are known from Ref.~\cite{BFP98}.

Let us start by deriving a few lowest-order results. First, it is
immediate to see that the non-spin parts of the matter source densities
$\sigma$, $\sigma_i$ and $\sigma_{ij}$, Eqs.~\eqref{sigmamunu}, start at
Newtonian order, and that their spin parts start at 0.5PN order $\sim
1/c$ in the cases of the vectorial $\sigma^\text{S}_i$ and tensorial
densities $\sigma^\text{S}_{ij}$, and only at 1.5PN order $\sim 1/c^3$
in the case of the scalar density $\sigma^\text{S}$. Here we are using
our counting for the PN order of spins [see Eq.~\eqref{S}], which is
physically appropriate to maximally rotating compact objects. With
lowest-order precision the expressions of the source densities for two
spinning particles read
\begin{subequations}\label{sigmaSmunu}
\begin{align}
\mathop{\sigma}_\text{S} &= -
\frac{2}{c^3}\,\varepsilon_{ijk}\,v_1^i\,S_1^j\,\partial_k\delta_1+
1\leftrightarrow 2+\mathcal{O}\left(\frac{1}{c^5}\right) \,
,\label{sigmaS}\\ \mathop{\sigma}_\text{S}{}_{\!i} &= -
\frac{1}{2c}\,\varepsilon_{ijk}\,S_1^j\,\partial_k\delta_1+1\leftrightarrow
2+\mathcal{O}\left(\frac{1}{c^3}\right) \, , \\
\mathop{\sigma}_\text{S}{}_{\!ij} &= -
\frac{1}{c}\,\varepsilon_{kl(i}\,v_1^{j)}\,S_1^k\partial_l\delta_1+1\leftrightarrow
2+\mathcal{O}\left(\frac{1}{c^3}\right) \, .
\end{align}
\end{subequations}
The symbol $1\leftrightarrow 2$ means adding the same terms but
corresponding to the other particle. The Dirac delta-function is denoted
by $\delta_1\equiv\delta (\mathbf{x}-\mathbf{y}_1)$, and
$\partial_k\delta_1$ means the spatial gradient of $\delta_1$ with
respect to the field point $\mathbf{x}$. The lowest-order potentials are
then straightforward to obtain from the fact that
$\Delta(1/r_1)=-4\pi\,\delta_1$ (where
$r_1\equiv\vert\mathbf{x}-\mathbf{y}_1\vert$), and we get
\begin{subequations}\label{VS}
\begin{align}
\mathop{V}_\text{S} &= -
\frac{2G}{c^3}\,\varepsilon_{ijk}\,v_1^i\,S_1^j\,\partial_k
\left(\frac{1}{r_1}\right) +1\leftrightarrow
2+\mathcal{O}\left(\frac{1}{c^{5}}\right) \, , \\
\mathop{V}_\text{S}{}_{\!i} &= -
\frac{G}{2c}\,\varepsilon_{ijk}\,S_1^j\,\partial_k\left(\frac{1}{r_1}
\right)+1\leftrightarrow 2+\mathcal{O}\left(\frac{1}{c^{3}}\right) \,
,\label{ViS} \\ \mathop{\hat{W}}_\text{S}{}_{\!ij} &= - \frac{G}{c}
\,\varepsilon_{kl(i}\,v_1^{j)}\,S_1^k\,\partial_l\left(\frac{1}{r_1}\right)+
\frac{G}{c}\,\delta_{ij}\,\varepsilon_{klm}\,v_1^k\,S_1^l\,
\partial_m\left(\frac{1}{r_1}\right) +1\leftrightarrow
2+\mathcal{O}\left(\frac{1}{c^{3}}\right) \, ,\label{WijS}\\
\mathop{\hat{W}}_\text{S}{}_{\!kk} &= \frac{2G}{c}
\,\varepsilon_{klm}\,v_1^k\,S_1^l\,\partial_m\left(\frac{1}{r_1}\right)
+1\leftrightarrow 2+\mathcal{O}\left(\frac{1}{c^{3}}\right) \, .
\end{align}
\end{subequations}
At the dominant level only contribute to the potentials some
compact-support terms (proportional to the source densities
$ \sigma^\text{S}_{\mu\nu} $) --- notably the non-compact support term
$\sim\partial V\partial V$ in the spin part of the potential
$\hat{W}_{ij}$, Eq.~\eqref{Wij}, turns out to be negligible.

To find all the spin terms in the equations of motion up to 2.5PN order,
we see from Eq.~\eqref{exprPAFA} that we need $V$ to 2.5PN order and
$V_i$ at 1.5PN order [\textit{i.e.} 1PN beyond what is given by
\eqref{ViS}], together with $\hat{W}_{ij}$, $\hat{R}_{i}$ and $\hat{X}$
at order 0.5PN. As we see, the potential $\hat{W}_{ij}$ is already given
by Eq.~\eqref{WijS} with the right precision. Our first problem is to
obtain the compact-support ``Newtonian'' potential $V$ to the 2.5PN
order. Definition~\eqref{V} shows that the mass density $ \sigma $,
source of $ V $, admits at an arbitrary high PN order the structure
\begin{equation}\label{sigmastruct}
\sigma = \Big( \tilde{\mu}_1 + \mathop{\tilde{\mu}_1}_\text{S} \Big)
\delta_1 + \frac{1}{\sqrt{-g}} \, \partial_t \Big(
\mathop{\nu_1}_\text{S} \delta_1 \Big) + \frac{1}{\sqrt{-g}} \,
\partial_i \Big( \mathop{\nu^i_1}_\text{S} \delta_1 \Big) + 1
\leftrightarrow 2 \, .
\end{equation}
The factors $ \tilde{\mu}_A $, ${}_\text{S}\tilde{\mu}_A$, ${}_\text{S}
\nu_A$ and ${}_\text{S} \nu^i_A$ are functions of the spins and the
velocities $v_A^i$, and functionals of the metric components or,
equivalently at 2.5PN, of the elementary potentials~\eqref{potentials}.
Note that though $ g_{\mu\nu}(\mathbf{x},t) $, by contrast to $
S^{\mu\nu}_A(t) $, depends on the field point, this is not the case of
the moment-like quantities entering the square brackets of
Eq.~\eqref{TmunuS3}. Each of them, being multiplied by the Dirac
distributions $\delta_A$, is indeed evaluated at point $\mathbf{x} =
\mathbf{y}_A$, after the Hadamard procedure described below. Thus, it
depends on time only (\textit{via} the point-mass positions
$\mathbf{y}_A$ and velocities $ \mathbf{v}_A $). The index S indicates
an additional linear dependence in the spin components, but of course,
the full spin dependence is more complicated due to the implicit
occurrence of $ S^{\mu\nu}_A $ in the potentials themselves. Notably,
the effective mass $ \tilde{\mu}_1 $ whose expression in terms of $V$,
$V_i$, $\hat{W}_{ii}$ and $v_1^2$ can be found in Ref.~\cite{BFP98}
contains a net contribution due to the spin at the 2.5PN order and given
by
\begin{equation}\label{mu1S}
\big(\tilde{\mu}_1\big)_\text{S} = m_1 \biggl( - \frac{1}{c^2}
\mathop{V}_\text{S} + \frac{1}{c^4} \Bigl[ -4 \mathop{V_i}_\text{S}
v_1^i - 2 \mathop{\hat{W}_{ii}}_\text{S} \Bigr] \biggr)_1 + \mathcal{O}
\left(\frac{1}{c^6}\right) \, ,
\end{equation}
where the value at the particle's location is meant in the sense of
Eq.~\eqref{hadamarda}. The expressions of the other moments will not be
provided here. It is in fact sufficient for our purpose to observe that,
as shown by Eq.~\eqref{mu1S}, we have $ (\tilde{\mu}_1)_\text{S} +
{}_\text{S}\tilde{\mu}_1 =
\mathcal{O}(1/c^5)$, and that ${}_\text{S}\nu_1 $ is at least of order $
\mathcal{O}(1/c^7)$ whereas ${}_\text{S}\nu^i_1$ is of order
$\mathcal{O}(1/c^3)$.

As the spin contribution in $ \sigma $, say $ {}_\text{S}\sigma$, is
already of order 1.5PN $\sim 1/c^3$, see Eq.~\eqref{sigmaS}, we need to
expand the retardations in $V$ only at relative 1PN order, hence
\begin{align}\label{VSexp}
\mathop{V}_\text{S} &= G
\int\frac{d^3\mathbf{x}'}{\vert\mathbf{x}-\mathbf{x}'\vert}\, \Big(
\sigma(\mathbf{x}',t) \Big)_\text{S} - \frac{G}{c} \int
d^3\mathbf{x}'\,\bigg( \frac{\partial}{\partial t}\sigma(\mathbf{x}',t)
\bigg)_\text{S}\nonumber\\ &+ \frac{G}{2 c^2}\int
d^3\mathbf{x}'\,\vert\mathbf{x}-\mathbf{x}'\vert\, \bigg(
\frac{\partial^2}{\partial t^2}\sigma(\mathbf{x}',t) \bigg)_\text{S}
+\mathcal{O}\left(\frac{1}{c^{5}}\right) \, .
\end{align}
We then substitute the value of $ \sigma $ following from
Eq.~\eqref{sigmastruct}. The integrals are evaluated with the help of
the formulas
\begin{subequations}\label{hadamard}
\begin{align}\label{hadamarda}
 \int d^3 \mathbf{x}' \,F(\mathbf{x}') \, \delta(\mathbf{x}' -
\mathbf{y}_1) &= (F)_1 \, , \\\int d^3 \mathbf{x}' \,F(\mathbf{x}') \,
\partial'_i \delta(\mathbf{x}' - \mathbf{y}_1) &= - (\partial_i F)_1 \,,
\end{align}
\end{subequations}
where the values at point $\mathbf{y}_1$ are denoted by parenthesis like
for $(F)_1$. These formulas extend the usual formulas of distribution
theory, which are valid for a smooth function $F$ with compact support,
to singular functions with a finite number of singular points and
deprived of essential singularities (see Ref.~\cite{BFreg} for full
explanations about this generalization). The formulas~\eqref{hadamard}
are part of Hadamard's self-field regularization which is systematically
employed in the present approach and the one
of~\cite{BFeom, BF00}.\,\footnote{Hadamard's regularization is known to yield
some ambiguous coefficients in the equations of motion and the radiation
field of non-spinning point particles at 3PN order. When using
dimensional regularization these ambiguities are seen to be associated
with the appearance of poles $\propto 1/\varepsilon$ (or ``cancelled''
poles) in the dimension of space $d=3+\varepsilon$~\cite{BDEI04}. The PN
order considered in the present paper is merely 1PN, since we are
computing the 1PN correction to the leading spin-orbit effect. At this
order there are no poles; therefore dimensional and Hadamard's
regularizations are equivalent.} In the end we are led to
\begin{align}\label{Vfinal}
  \mathop{V}_\text{S}(\mathbf{x},t) &= \frac{G}{r_1}\left[
  \mathop{\tilde{\mu}_1}_\text{S}(t) +
  \big(\tilde{\mu}_1\big)_\text{S}(t)\right] - G
  \mathop{\nu_1^i}_\text{S}(t)\,\biggl( \partial'_i \Bigl(
  \frac{1}{\sqrt{-g(\mathbf{x}',t)} \, |\mathbf{x} - \mathbf{x}'|}
  \Bigr) \biggr)_1 \nonumber \\ & + \frac{G}{2 c^2}\,\partial_t^2 \Bigl(
  \mathop{\nu_1^i}_\text{S}\partial_i r_1 \Bigr) - \frac{G}{2
  c^2}\,m_1\big(a_1^i\big)_\text{S} \partial_i r_1 + 1 \leftrightarrow 2
  + \mathcal{O} \left(\frac{1}{c^6} \right) \, .
\end{align}
The final result for $ V $ is obtained by replacing the moments and the
determinant of the metric at the 2.5PN level by their explicit values
derived from the lowest order approximation of the potentials. The
computation of ${}_\text{S}V_i$ is similar to that of $ {}_\text{S}V $,
though slightly simpler since the counterpart of $ \mathop{\tilde{\mu}}
$ for $ \sigma_i $ does not depend implicitly on the spin at the 1.5PN
order.

Next we explain how to compute the non-compact (NC) support terms, and
we take the example of the particular NC term in the potential
$\hat{R}_{i}$ given by
\begin{equation}\label{RiSNC}
\mathop{\hat{R}}_\text{S}{}_{\!i}^\mathrm{(NC)} =
\Delta^{-1}\left[-2\partial_{k}V\partial_{i}\mathop{V}_\text{S}{}_{\!k}\right]
+ \mathcal{O}\left(\frac{1}{c^{3}}\right)\,.
\end{equation}
In the source of this term we have to insert the Newtonian approximation
of the potential $V$, which is simply $
V=\frac{G\,m_1}{r_1}+\frac{G\,m_2}{r_2}+\mathcal{O}(c^{-2}) $, together
with the leading-order spin term ${}_\text{S}V_k$ given previously in
Eq.~\eqref{ViS}. The source being known we are then able to integrate
(using the same techniques as in Ref.~\cite{BFP98}) and we get
\begin{equation}\label{RiSNCresult}
\mathop{\hat{R}}_\text{S}{}_{\!i}^\mathrm{(NC)} = \frac{G^2 m_1}{8
c}\,\varepsilon_{ikl}\,S_1^k\partial_{l}\left(\frac{1}{r_1^2}\right) -
\frac{G^2
m_2}{c}\,\varepsilon_{klm}\,S_1^k\,\mathop{\partial}_1{}_{\!il}
\mathop{\partial}_2{}_{\!m} g + 1\leftrightarrow
2+\mathcal{O}\left(\frac{1}{c^{3}}\right) \, ,
\end{equation}
in which
\begin{subequations}\label{g}
\begin{equation}
g = \ln \left(r_1 + r_2 + r_{12}\right)
\end{equation}
satisfies
\begin{equation}
\Delta g = \frac{1}{r_1\,r_2} \, . 
\end{equation}
\end{subequations}
The crucial fact which enables the latter integration in closed analytic
form is the existence of the function $g$ (first introduced by Fock
\cite{Fock}). This function and its generalizations are extremely useful
in the computation of the spinless equations of motion at 2PN and 3PN
orders~\cite{BFP98, BFeom}.

Finally all the necessary spin parts of the potentials are computed by
PN iteration, ready for insertion into the non-spin contribution of the
equations of motion as given by Eqs.~\eqref{exprPAFA}. For all the
potentials we are in agreement with the results reported by Tagoshi
\textit{et al.}~\cite{TOO01} in their Appendix.\,\footnote{We have
however noticed the following misprints in Ref.~\cite{TOO01}: in
Eq.~(A1h) for $ {}_\text{S}\hat{R}_i$, the third term in the first
parenthesis of the first line should be $ + m_2/(r_{12} s^2) $; in
Eq.~(A1i) for $ {}_\text{S}\hat{X} $, the first term in the parenthesis
following $ (n_{12}v_2) $ in the third line must be read $ -m_2/(r_{12}
s^2) $.}

\section{The 2.5PN equations of motion with spin-orbit effects}\label{secV}

\subsection{Equations in a general frame}

In addition to the spin parts of the potentials computed in
Sec.~\ref{secIV} and inserted into Eqs.~\eqref{exprPAFA}, we add the
required spin corrections to the geodesic motion as given by either the
formulation of Eqs.~\eqref{PAFAspin} or that
of~\eqref{PAFApapa}--\eqref{PAFAresult}. The latter corrections are
computed by inserting into them the non-spin parts of the potentials
taken from~\cite{BFP98}. We find that the two formulations [respectively
given by~\eqref{PAFAspin} and \eqref{PAFApapa}--\eqref{PAFAresult}] are
equivalent and agree on the result. Finally the 2.5PN equations of
motion with spin-orbit effects are obtained in the form
\begin{equation}\label{a1struct}
\frac{d \mathbf{v}_1}{d t}=
\mathbf{A}_\mathrm{N}+\frac{1}{c^2}\mathbf{A}_\mathrm{1PN}+\frac{1}{c^3}
\mathop{\mathbf{A}}_\text{S}{}_{\!\mathrm{1.5PN}}
+\frac{1}{c^4}\left[\mathbf{A}_\mathrm{2PN}+
\mathop{\mathbf{A}}_\text{SS}{}_{\!\mathrm{2PN}}\right]
+\frac{1}{c^5}\left[\mathbf{A}_\mathrm{2.5PN}+
\mathop{\mathbf{A}}_\text{S}{}_{\!\mathrm{2.5PN}}\right]
+\mathcal{O}\left(\frac{1}{c^6}\right) \, .
\end{equation}
Here the Newtonian acceleration is
$A^i_\mathrm{N}=-\frac{G\,m_2}{r_{12}^2}\,n_{12}^i$, and we denote by
$A^i_\mathrm{N}$, $A^i_\mathrm{1PN}$, $A^i_\mathrm{2PN}$ and
$A^i_\mathrm{2.5PN}$ the standard non-spin contributions (in harmonic
coordinates) which are well-known, see Eqs.~(8.4) in~\cite{BFP98} and
earlier works reviewed in~\cite{D83houches}. In particular
$A^i_\mathrm{2.5PN}$ represents the standard radiation reaction damping
term. (For simplicity we henceforth suppress the subscript NS on
non-spin-type contributions.)

The leading-order spin effect is the 1.5PN spin-orbit term. For this
term we recover the standard expression, known from
Refs.~\cite{BOC75, BOC79} and given in~\cite{KWWi93, K95} in the
center-of-mass frame, and in~\cite{TOO01} in a general frame. In the
following we shall sometimes use some formulas relating the ``mixed
products'' of three vectors in three dimensions,
\begin{subequations}\label{eq:dependence}
\begin{align} 
(U_1, U_2, U_3) \, \mathbf{U} &= (UU_1) \, \mathbf{U}_2 \times
\mathbf{U}_3 + (UU_2) \, \mathbf{U}_3 \times \mathbf{U}_1 + (UU_3) \,
\mathbf{U}_1 \times \mathbf{U}_2 \\&= (U, U_2, U_3) \, \mathbf{U}_1 +
(U_1, U, U_3) \, \mathbf{U}_2 + (U_1, U_2, U) \, \mathbf{U}_3\, ,
\end{align}\end{subequations}
valid for any vectors $ \mathbf{U}$, $ \mathbf{U}_1 $, $ \mathbf{U}_2 $,
$ \mathbf{U}_3 $ (in 3 dimensions). Here the vectorial product of
ordinary Euclidean vectors is indicated with the $\times$ symbol, for
instance $(\mathbf{U}_1 \times
\mathbf{U}_2)^i=\varepsilon^{ijk}U_1^jU_2^k$; parenthesis denote the
usual Euclidean scalar product,
$(UU_1)=U^iU_1^i=\mathbf{U}\cdot\mathbf{U}_1$; and the mixed product, or
determinant between three vectors, is denoted $(U_1, U_2, U_3)\equiv
\mathbf{U}_1\cdot(\mathbf{U}_2 \times
\mathbf{U}_3)=\varepsilon_{ijk}U_1^iU_2^jU_3^k$. This yields
\begin{subequations}
\begin{align}\label{A1.5PN}
\mathop{\mathbf{A}}_\text{S}{}_{\!\mathrm{1.5PN}}= \frac{G
m_2}{r_{12}^3}& \bigg\{ \bigg[6 \frac{(S_1, n_{12}, v_{12})}{m_1} + 6
\frac{(S_2, n_{12}, v_{12})}{m_2} \bigg] \mathbf{n}_{12} + 3
(n_{12}v_{12}) \frac{\mathbf{n}_{12} \times \mathbf{S}_1}{m_1} \nonumber
\\ & + 6 (n_{12}v_{12}) \frac{\mathbf{n}_{12} \times \mathbf{S}_2}{m_2}
- 3 \frac{\mathbf{v}_{12} \times \mathbf{S}_1}{m_1} - 4
\frac{\mathbf{v}_{12} \times \mathbf{S}_2}{m_2} \bigg\} \,.
\end{align}
We use, whenever convenient, the notation
$\mathbf{v}_{12}=\mathbf{v}_{1}-\mathbf{v}_{2}$ for the relative
velocity.

The next-order spin correction is the spin-spin (SS) at 2PN order. We do
not give this term since we are concerned here with spin-orbit effects
which are linear in the spins. The SS term is quadratic in the spins,
$\mathcal{O}(S^2)$, and can be found in Refs.~\cite{BOC75, BOC79} and
\textit{e.g.} in Eq.~(5.9) of~\cite{TOO01}. Now the 1PN correction to
the spin-orbit effect, which is the aim of this paper and the work
\cite{TOO01}, reads
\begin{align}\label{A2.5PN}
& \mathop{\mathbf{A}}_\text{S}{}_{\!\mathrm{2.5PN}} = \frac{G
 m_2}{r_{12}^3} \bigg\{ \mathbf{n}_{12} \bigg[- 6 (n_{12}, v_1, v_2)
 \bigg( \frac{(v_1S_1)}{m_1} + \frac{(v_2S_2)}{m_2} \bigg) \nonumber \\
 & \qquad\qquad -\frac{(S_1, n_{12}, v_{12})}{m_1} \bigg(15
 (n_{12}v_2)^2 + 6 (v_{12}v_2) + 26 \frac{G m_1}{r_{12}} + 18 \frac{G
 m_2}{r_{12}} \bigg) \nonumber \\ & \qquad\qquad - \frac{(S_2, n_{12},
 v_{12})}{m_2} \bigg(15 (n_{12}v_2)^2 + 6 (v_{12}v_2) + \frac{49}{2}
 \frac{G m_1}{r_{12}}+ 20 \frac{G m_2}{r_{12}} \bigg) \bigg] \nonumber
 \\ & \qquad + \mathbf{v}_1 \bigg[-3 \frac{(S_1, n_{12}, v_1)}{m_1}
 \bigg( (n_{12}v_1) + (n_{12}v_2) \bigg) + 6 (n_{12}v_1) \frac{(S_1,
 n_{12}, v_2)}{m_1} - 3 \frac{(S_1, v_1, v_2)}{m_1} \nonumber \\ &
 \qquad\qquad - 6 (n_{12}v_1) \frac{(S_2, n_{12}, v_1)}{m_2} +
 \frac{(S_2, n_{12}, v_2)}{m_2} \bigg( 12 (n_{12}v_1) - 6 (n_{12}v_2)
 \bigg) - 4 \frac{(S_2, v_1, v_2)}{m_2}\bigg] \nonumber \\ & \qquad +
 \mathbf{v}_2 \bigg[6 (n_{12}v_1) \frac{(S_1, n_{12},v_{12})}{m_1} + 6
 (n_{12}v_1) \frac{(S_2, n_{12}, v_{12})}{m_2} \bigg] \nonumber \\ &
 \qquad - \mathbf{n}_{12} \times \mathbf{v}_1 \bigg[3 (n_{12}v_{12})
 \frac{(v_1S_1)}{m_1}+ 4 \frac{G m_1}{r_{12}} \frac{(n_{12}S_2)}{m_2}
 \bigg] \nonumber \\ & \qquad - \mathbf{n}_{12} \times \mathbf{v}_2
 \bigg[6 (n_{12}v_{12}) \frac{(v_2S_2)}{m_2} - 4 \frac{G m_1}{r_{12}}
 \frac{(n_{12}S_2)}{m_2} \bigg] + \mathbf{v}_1 \times \mathbf{v}_2
 \bigg[3 \frac{(v_1S_1)}{m_1} + 4 \frac{(v_2S_2)}{m_2}\bigg] \nonumber
 \\ & \qquad + \frac{\mathbf{n}_{12} \times \mathbf{S}_1}{m_1}
 \bigg[-\frac{15}{2} (n_{12}v_{12}) (n_{12}v_2)^2 + 3 (n_{12}v_2)
 (v_{12}v_2) \nonumber \\ & \qquad\qquad - 14 \frac{G m_1}{r_{12}}
 (n_{12}v_{12}) -9 \frac{G m_2}{r_{12}} (n_{12}v_{12}) \bigg] \nonumber
 \\ & \qquad + \frac{\mathbf{n}_{12} \times \mathbf{S}_2}{m_2} \bigg[-15
 (n_{12}v_{12}) (n_{12}v_2)^2 - 6 (n_{12}v_1) (v_{12}v_2) + 12
 (n_{12}v_2) (v_{12}v_2) \nonumber \\ & \qquad\qquad + \frac{G
 m_1}{r_{12}} \bigg(- \frac{35}{2} (n_{12}v_1) + \frac{39}{2}
 (n_{12}v_2) \bigg)- 16 \frac{G m_2}{r_{12}} (n_{12}v_{12}) \bigg]
 \nonumber \\ & \qquad + \frac{\mathbf{v}_{12} \times \mathbf{S}_1}{m_1}
 \bigg[-3 (n_{12}v_1) (n_{12}v_2) + \frac{15}{2} (n_{12}v_2)^2 +
 \frac{G}{r_{12}} (14 m_1 + 9 m_2) + 3 (v_{12}v_2) \bigg] \nonumber \\ &
 \qquad + \frac{\mathbf{v}_{12} \times \mathbf{S}_2}{m_2} \bigg[6
 (n_{12}v_2)^2 + 4 (v_{12}v_2) + \frac{23}{2} \frac{G m_1}{r_{12}} + 12
 \frac{G m_2}{r_{12}} \bigg] \bigg\} \, .
\end{align}
\end{subequations}
Surprisingly, we find that our expression has substantial differences
with the result given in Eq.~(5.10) of~\cite{TOO01}. However, since we
recovered in the last section exactly the same potentials as given in
the Appendix of \cite{TOO01}, and since as we shall see below we find
perfect agreement with the equations of motion computed in~\cite{TOO01}
in the case of the center-of-mass frame, we believe that the latter
differences can only be due to some trivial misprints (and most probably
to some mixup of Mathematica files) in the last stage of the
work~\cite{TOO01}.\,\footnote{For completeness we indicate here all the
misprints in Eq.~(5.10) of~\cite{TOO01}: in order to recover the correct
acceleration, the last term before the closing curly brackets on the
fifth line of Eq.~(5.10), $-12 \varepsilon^{jkl} n^j v_1^k v_2^l
(v_1S_1)/m_1 $, must be replaced by $ -6 \varepsilon^{jkl} n^j v_1^k
v_2^l [(v_1S_1)/m_1 + (v_2S_2)/m_2] $; the term before the last one in
the seventh line has to be read $-6 n^j v_2^k (nv_{12}) (v_2S_2)/m_2 $
instead of $-6 n^j v_2^k (nv_{12}) (v_1S_1)/m_1 $; and the very last
term $+7 v_1^j v_2^k (v_1S_1)/m_1 $ must be modified as $ + v_1^j v_2^k
[3 (v_1S_1)/m_1 + 4 (v_2S_2)/m_2] $.}

In Appendix~\ref{secA} we shall prove that the equations of motion stay
invariant under global Poincar\'e transformations. Such a verification
is quite important to test the correctness of the equations (it played
an important role during the computation of the 3PN non-spin terms in
\cite{BFeom, BF00}). Furthermore, we show in Appendix~\ref{secB} that the test
mass limit of the equations of motion is identical with the geodesic
equations around a Kerr black hole (for simplicity we restrict ourself
to circular orbits). Both verifications have already been made in
Ref.~\cite{TOO01} but we present some alternative ways to do the checks.

\subsection{Equations in the center-of-mass frame}

Let us now present the result in the center-of-mass (CM) frame, defined
by the nullity of the center-of-mass vector, equal to the conserved
integral associated with the boost invariance of the equations of
motion, which will be checked in Appendix~\ref{secA}. We shall derive
the center-of-mass integral at the 2.5PN order in the next section,
however for the present computation we need it only at the 1.5PN order.
When working in the CM frame we find it convenient to introduce the same
spin variables as chosen by Kidder~\cite{K95} (except that we denote by
$\mathbf{\Sigma}$ what he calls $\mathbf{\Delta}$), namely
\begin{subequations}\label{SSigma}
\begin{align}
\mathbf{S} &\equiv \mathbf{S}_1 + \mathbf{S}_2 \, , \\ \mathbf{\Sigma}
&\equiv m \Big(\frac{\mathbf{S}_2}{m_2} - \frac{\mathbf{S}_1}{m_1}
\Big)\, .
\end{align}
\end{subequations}
Mass parameters are denoted by $m\equiv m_1+m_2$, $\delta m \equiv
m_1-m_2$ and $\nu\equiv m_1\,m_2/m^2$ (such that $0<\nu\leq 1/4$). At
the leading order in the spins we have the following relation between
the positions $\mathbf{y}_1$ and $\mathbf{y}_2$ in the CM frame and the
relative position $\mathbf{x}=\mathbf{y}_1-\mathbf{y}_2$ and velocity
$\mathbf{v}=d\mathbf{x}/d t=\mathbf{v}_1-\mathbf{v}_2=\mathbf{v}_{12}$
(see \textit{e.g.} Ref.~\cite{TOO01})
\begin{subequations}\label{y12CM}
\begin{align}
\mathbf{y}_1 &= \left[\frac{m_2}{m} + \frac{\nu}{2 c^2}\frac{\delta
m}{m}\left(v^2-\frac{G\,m}{r}\right)\right]\,\mathbf{x} +
\frac{\nu}{m\,c^3}\,\mathbf{v} \times \mathbf{\Sigma} \, , \\
\mathbf{y}_2 &= \left[-\frac{m_1}{m} + \frac{\nu}{2 c^2}\frac{\delta
m}{m}\left(v^2-\frac{G\,m}{r}\right)\right]\,\mathbf{x} +
\frac{\nu}{m\,c^3}\,\mathbf{v}\times \mathbf{\Sigma} \, .
\end{align}
\end{subequations}
In addition to the spin-orbit effect at order 1.5PN $\sim 1/c^3$ (last term in
these relations), we have included the well-known 1PN $\sim 1/c^2$
non-spin term. This term is obviously needed here because during the
reduction of the equations of motion to the CM frame at order 2.5PN in
the spins, we shall need to take into account the 1PN non-spin term
coupled to the lowest-order 1.5PN spin term; such coupling evidently produces
some 2.5PN spin terms. In the CM frame the equation of the relative
motion reads
\begin{equation}\label{aCM}
\frac{d \mathbf{v}}{d t}=
\mathbf{B}_\mathrm{N}+\frac{1}{c^2}\mathbf{B}_\mathrm{1PN}+\frac{1}{c^3}
\mathop{\mathbf{B}}_\text{S}{}_{\!\mathrm{1.5PN}}
+\frac{1}{c^4}\left[\mathbf{B}_\mathrm{2PN}+
\mathop{\mathbf{B}}_\text{SS}{}_{\!\mathrm{2PN}}\right]
+\frac{1}{c^5}\left[\mathbf{B}_\mathrm{2.5PN}+
\mathop{\mathbf{B}}_\text{S}{}_{\!\mathrm{2.5PN}}\right]
+\mathcal{O}\left(\frac{1}{c^6}\right) \, ,
\end{equation}
where we recognize all the various terms similarly to
Eq.~\eqref{a1struct}. We find that the spin-orbit term and the 1PN correction 
to the spin-orbit are given, in terms of the spin
variables~\eqref{SSigma}, by
\begin{subequations}\label{BSO}\begin{align}
\mathop{\mathbf{B}}_\text{S}{}_{\!\mathrm{1.5PN}} = \frac{G}{r^3} &\bigg\{
\mathbf{n} \bigg[12 (S, n, v) + 6 \frac{\delta m}{m} (\Sigma, n, v) \bigg] + 
9 (nv) \mathbf{n} \times \mathbf{S} + 
3 \frac{\delta m}{m} (nv) \mathbf{n} \times \mathbf{\Sigma} \nonumber \\ & - 7
\mathbf{v} \times \mathbf{S} -  
   3 \frac{\delta m}{m} \mathbf{v} \times \mathbf{\Sigma} \bigg\} \, , \\
\mathop{\mathbf{B}}_\text{S}{}_{\!\mathrm{2.5PN}} = \frac{G}{r^3} & \bigg\{  
\mathbf{n} \bigg[(S, n, v) \bigg(-30 \nu (nv)^2  + 24 \nu v^2 - 
\frac{G m}{r} (38 + 25 \nu) \bigg) \nonumber \\ & \qquad + 
\frac{\delta m}{m} (\Sigma, n, v) \bigg(-15 \nu (nv)^2 + 
12 \nu v^2 - \frac{G m}{r} (18  + \frac{29}{2} \nu) \bigg) \bigg] \nonumber 
\\ & +  (nv) \mathbf{v} \bigg[ (S, n, v) (-9 + 9 \nu) + \frac{\delta m}{m}
(\Sigma, n, v) (-3 + 6 \nu) \bigg] \nonumber \\ & + 
\mathbf{n} \times \mathbf{v} \bigg[(nv) (vS) (-3 + 3 \nu) -
8 \frac{G m}{r} \nu (nS) - 
\frac{\delta m}{m} \bigg(4 \frac{G m}{r} \nu (n\Sigma) + 
3 (nv) (v\Sigma)  \bigg) \bigg] \nonumber \\ & + 
(nv) \mathbf{n} \times \mathbf{S} \bigg[-\frac{45}{2} \nu (nv)^2 + 
21 \nu v^2 - \frac{G m}{r} (25 + 15 \nu) \bigg] \nonumber \\ & +
 \frac{\delta m}{m} (nv) \mathbf{n} \times \mathbf{\Sigma} \bigg[-
15 \nu (nv)^2  + 12 \nu v^2 - \frac{G m}{r} (9 + \frac{17}{2} \nu)\bigg]
\nonumber \\ & + 
\mathbf{v} \times \mathbf{S}  \bigg[\frac{33}{2} \nu (nv)^2 + 
\frac{G m}{r} (21 + 9 \nu) - 14 \nu v^2 \bigg] \nonumber \\ & + 
\frac{\delta m}{m} \mathbf{v} \times \mathbf{\Sigma} \bigg[9 \nu (nv)^2 - 
7 \nu v^2 + \frac{G m}{r} (9 + \frac{9}{2} \nu) \bigg] \bigg\} \, .
\end{align}\end{subequations}
We find perfect agreement with Eqs.~(5.18) and (5.20) of Tagoshi
\textit{et al.}~\cite{TOO01}.\,\footnote{Note that the spin variables
adopted in~\cite{TOO01} are defined by $\boldsymbol{\chi}_s\equiv
\frac{1}{2}\left(\frac{\mathbf{S}_1}{m_1^2}+
\frac{\mathbf{S}_2}{m_2^2}\right)$ and $\boldsymbol{\chi}_a\equiv
\frac{1}{2}\left(\frac{\mathbf{S}_1}{m_1^2}-
\frac{\mathbf{S}_2}{m_2^2}\right)$ and differ from our own. We have
$$\mathbf{S}=m^2\Bigl[(1-2\nu)\boldsymbol{\chi}_s+\frac{\delta
m}{m}\boldsymbol{\chi}_a\Bigr] ~~\text{and}~~
\mathbf{\Sigma}=m^2\Bigl[-\frac{\delta m}{m}\boldsymbol{\chi}_s
-\boldsymbol{\chi}_a\Bigr] \, .$$ }

\subsection{Reduction to quasi circular orbits}

Finally we present the case where the orbit is nearly circular,
\textit{i.e.} whose radius is constant apart from small perturbations
induced by the spins (as usual we neglect the gravitational radiation
damping at 2.5PN order). Following Ref.~\cite{K95} we introduce an
orthonormal triad $\{\mathbf{n},\bm{\lambda},\bm{\ell}\}$ defined by
$\mathbf{n}=\mathbf{x}/r$ as before,
$\bm{\ell}=\mathbf{L}_\mathrm{N}/\vert\mathbf{L}_\mathrm{N}\vert$ where
$\mathbf{L}_\mathrm{N}\equiv\mu\,\mathbf{x}\times\mathbf{v}$ denotes the
Newtonian angular momentum, and
$\bm{\lambda}=\bm{\ell}\times\mathbf{n}$. The orbital frequency $\omega$
is defined for general, not necessarily circular orbits,
$\mathbf{v}=\dot{r}\mathbf{n}+r\omega\bm{\lambda}$ where $\dot{r}=(nv)$.
The components of the acceleration $\mathbf{a}=d\mathbf{v}/dt$ along the
basis $\{\mathbf{n},\bm{\lambda},\bm{\ell}\}$ are then given by
\begin{subequations}\label{acccomp}\begin{align}
\mathbf{n}\cdot\mathbf{a} &= \ddot{r} - r \omega^2\,,\\
\bm{\lambda}\cdot\mathbf{a} &= r \dot{\omega} + 2 \dot{r} \omega\,,\\
\bm{\ell}\cdot\mathbf{a} &= - r \omega \Bigl(\bm{\lambda}\cdot\frac{d
\bm{\ell}}{d t}\Bigr)\,.
\end{align}\end{subequations}
We project out the spins on this orthonormal basis, defining
$\mathbf{S}= S_n \mathbf{n} + S_\lambda \bm{\lambda} + S_\ell \bm{\ell}$
and similarly for $\mathbf{\Sigma}$. Next we impose the restriction to
circular orbits which means $\ddot{r}=0=\dot{r}$ and $v^2 = r^2\omega^2$
(neglecting radiation reaction damping terms). In this way we find that
the equations of motion~\eqref{aCM} with~\eqref{BSO} are of the type
\begin{equation}\label{acirc}
\frac{d \mathbf{v}}{d t} = - \omega^2 r\,\mathbf{n} + a_\ell\,\bm{\ell}
+ \mathcal{O}\left(\frac{1}{c^6}\right) \, .
\end{equation}
There is no component of the acceleration along $\bm{\lambda}$.
Comparing with Eqs.~\eqref{acccomp} in the case of circular orbits, we
see that $\omega$ is indeed the orbital frequency, while $a_\ell=- r
\omega (\bm{\lambda}\cdot d \bm{\ell}/d t)$ is proportional to the
variation of $\bm{\ell}$ in the direction of the velocity
$\mathbf{v}=r\omega\bm{\lambda}$. We find that $\omega^2$ is of the form
\begin{equation}\label{omega} 
\omega^2 = \frac{G\,m}{r^3}\biggl\{1 + \frac{1}{c^2}\zeta_\mathrm{1PN}
+\frac{1}{c^3}\mathop{\zeta}_\text{S}{}_{\!\mathrm{1.5PN}} +
\frac{1}{c^4}\left[\zeta_\mathrm{2PN}
+\mathop{\zeta}_\text{SS}{}_{\!\mathrm{2PN}}\right]+\frac{1}{c^5}
\mathop{\zeta}_\text{S}{}_{\!\mathrm{2.5PN}}\biggr\}
+\mathcal{O}\left(\frac{1}{c^6}\right)\,,
\end{equation}
where $\zeta_\mathrm{1PN}$ and $\zeta_\mathrm{2PN}$ denote the standard
non-spin contributions,\,\footnote{They are given by
$\zeta_\mathrm{1PN}=\frac{G m}{r}\left(-3+\nu\right)$ and
$\zeta_\mathrm{2PN}=\left(\frac{G
m}{r}\right)^2\left(6+\frac{41}{4}\nu+\nu^2\right)$ in harmonic
coordinates.} and where
\begin{subequations}\label{zeta}\begin{align} 
\mathop{\zeta}_\text{S}{}_{\!\mathrm{1.5PN}} &=
\left(\frac{G\,m}{r}\right)^{3/2}\!\frac{1}{G\,m^2}\left[-5S_\ell - 3
\frac{\delta m}{m}\,\Sigma_\ell\right]\,,\\
\mathop{\zeta}_\text{S}{}_{\!\mathrm{2.5PN}} &=
\left(\frac{G\,m}{r}\right)^{5/2}\!\frac{1}{G\,m^2}\left[\left(\frac{39}{2}
-\frac{23}{2}\nu\right) S_\ell + \left(\frac{21}{2}
-\frac{11}{2}\nu\right)\frac{\delta m}{m}\,\Sigma_\ell\right]\,,
\end{align}\end{subequations}
with, \textit{e.g.} $S_\ell\equiv (S\ell) = \mathbf{S}\cdot\bm{\ell}$.
On the other hand, we get
\begin{equation}\label{aell} 
a_\ell = \frac{1}{c^3}\mathop{\alpha}_\text{S}{}_{\!\mathrm{1.5PN}} +
\frac{1}{c^4}\mathop{\alpha}_\text{SS}{}_{\!\mathrm{2PN}}+\frac{1}{c^5}
\mathop{\alpha}_\text{S}{}_{\!\mathrm{2.5PN}}+\mathcal{O}\left(\frac{1}{c^6}\right)\,,
\end{equation}
with spin-orbit coefficients
\begin{subequations}\label{alpha}\begin{align} 
\mathop{\alpha}_\text{S}{}_{\!\mathrm{1.5PN}} &=
\left(\frac{G\,m}{r}\right)^{3/2}\!\frac{1}{m \,r^2}\left[7S_n + 3
\frac{\delta m}{m}\,\Sigma_n\right]\,,\\
\mathop{\alpha}_\text{S}{}_{\!\mathrm{2.5PN}} &=
\left(\frac{G\,m}{r}\right)^{5/2}\!\frac{1}{m
\,r^2}\left[\left(-\frac{63}{2} +\frac{\nu}{2}\right) S_n -
\frac{27}{2}\frac{\delta m}{m}\,\Sigma_n\right]\,.
\end{align}\end{subequations}

We see that the resulting motion cannot be exactly circular for general
orientations of the spins. Let us show however that the time-averaged
acceleration coincides with the acceleration of a particle that rotates
uniformly about the origin. In a first step, we must make explicit the
time dependence of the dynamical variables $ \mathbf{x} $, $ \mathbf{S}
$ and $ \mathbf{\Sigma} $. As the motion is uniformly circular in the
absence of spin, the position $ \mathbf{x} $ decomposed along a fixed
orthonormal basis $\{\mathbf{e}_1, \mathbf{e}_2, \bm{\ell} \} $ reads
\begin{equation} \label{eq:x_components}
\mathbf{x}(t) = \mathbf{e}_1 r \cos
(\omega_\text{NS} t) + \mathbf{e}_2 r \sin (\omega_\text{NS} t) \, ,
\end{equation}
with $ \omega_\text{NS} $ being the orbital frequency when the spins are
turned off.

The spin variables are computed by means of the precession equations,
which decouple in the case of a pure spin-orbit interaction. The spin 1,
for instance, obeys an equation whose right-hand-side is polynomial in
$G m/r = v^2$, $ (nS_1) $ and $ (vS_1) $. For dimensional reasons, it
must then have the form (for circular orbits, up to say the 2PN order)
\begin{align} \label{eq:general_precession}
\frac{d\mathbf{S}_1}{dt} = \sum_{k=1,2} \Big(\frac{G m}{rc^2} \Big)^k & 
\bigg[a_{\text{S}_1}^{(k,n)} (vS_1)
\mathbf{n} + a_{\text{S}_1}^{(k,v)} (nS_1) \mathbf{v} \bigg] +
\mathcal{O}\bigg(\frac{1}{c^5}\bigg)  \, ,
\end{align}
and similarly for $ d\mathbf{S}_2/dt $. The functions of $m_1/m$,
$m_2/m$ denoted by $a_{\text{S}_1}^{(k,n)} $ and $a_{\text{S}_1}^{(k,v)}
$ may be obtained from the results of the next section (see also
paper~II). They allow us to define dimensionless coefficients like
$a^{(n)}_{\text{S}_1} = \sum_{k=1,2} (G m/r)^k a^{(k,n)}_{\text{S}_1}$.
The key point is that the latter coefficients are constant, which
suggests to solve the above differential equations in the moving basis $
\{\mathbf{n}, \bm{\lambda}, \bm{\ell} \}$. Indeed, the time derivative
of a spin component in this basis, say $ S^1_n = (nS_1)$, is given by a
relation of the type
\begin{equation}
\frac{dS^1_n}{dt} = (n\dot{S}_1) + (\dot{n}S_1)
\end{equation}
with $\dot{\mathbf{n}} = \dot{\mathbf{x}}/r = \omega_\text{NS}
\bm{\lambda} $. This results, after eliminating $\dot{\mathbf{S}}_1 $ by
means of Eq.~\eqref{eq:general_precession}, in a linear differential
equation with constant coefficients for $ S^1_n $. Proceeding in the
same way for the other components of the first spin, we arrive at the
following system:
\begin{equation} \label{eq:precession_system}
\frac{d\mathbf{X}_{S_1}}{dt} = \mathbf{M}_{S_1}. \mathbf{X}_{S_1} \,,
\end{equation}
where $ \mathbf{M}_{S_1} $ is a $3\times 3$ constant matrix and $
\mathbf{X}_{S_1} = (S^1_n,S^1_\lambda,S^1_\ell ) $. The relations $
\bm{\ell}. d\mathbf{S}_1/dt = \bm{\ell}. d\mathbf{S}_2/dt = 0$ (since
$\bm{\ell}$ is constant because we neglect the SS terms) imply that $
(0,0,1) $ is an eigenvector associated with the eigenvalue $ \lambda_0 =
0$. There remain two eigenvalues, say $ \lambda_1^+ $ and $ \lambda_1^-
$; but since the trace of $\mathbf{M}_{S_1} $ vanishes because $ (nv) =
0$, we have $ \lambda_1^-= -\lambda_1^+ $. In the end, we notice that
the spins are almost constant at Newtonian order in the basis $
\{\mathbf{e}_1,\mathbf{e}_2,\bm{\ell}\} $, which means that they precess
about $ \bm{\ell} $ with angular velocity $ -\omega_\text{NS} $ in the
moving frame. Therefore, $ \lambda_1^\pm $ is purely imaginary and
reduces to $\pm i \omega_\text{NS} $ at Newtonian order. At higher order
we shall have $ \lambda_1^\pm = \pm i (\omega_\text{NS} - \Omega_1) $
where $ \Omega_1 = \mathcal{O}(1/c^2)$ represents the precession
frequency. The components $S^1_n$ and $S^1_\lambda$ solving
Eq.~\eqref{eq:precession_system} are then linear combinations of $\cos
[(-\omega_\text{NS} + \Omega_1)t] $ and $\sin [(-\omega_\text{NS} +
\Omega_1)t] $.\,\footnote{This can also be deduced immediately from
introducing a different spin variable $\mathbf{S}_1^\text{c}$ with
constant magnitude (described in Sec. VII of paper~II) and obeying
$d\mathbf{S}_1^\text{c}/dt=\mathbf{\Omega}_1\times\mathbf{S}_1^\text{c}$;
noticing that the components of $\mathbf{S}_1^\text{c}$ in the basis $
\{\mathbf{n}, \bm{\lambda}, \bm{\ell} \}$ are linear combinations of
those of $\mathbf{S}_1$, with constant coefficients.} As for the
component $S^1_\ell$, it is constant neglecting terms quadratic in the
spins.

We complete our proof by time averaging the term $a_\ell$ in the
acceleration~\eqref{acirc}. We first observe that the conservative part
of the dynamics involves three different angular frequencies
($\omega_\text{NS}$, $\Omega_1$ and $\Omega_2$), so that it cannot be
periodic in general. Therefore, it is not appropriate to average the
particle motion on the orbital period. Instead, the time-average will be
achieved on infinite time. Defining
\begin{equation}
\langle S^1_n \rangle = \lim_{T \to + \infty} \frac{1}{T} \int_t^{t+T}
dt'~S^1_n(t') \, ,
\end{equation}
we find $\langle S^1_n \rangle=0$. We next notice that the orbital
frequency $ \omega $ is actually constant (neglecting SS terms), for it
depends on the spin through $ S^1_\ell $ and $S^2_\ell $ only, which are
constant. The average of $ a_\ell $ is a linear combination of $ \langle
S_n \rangle = \langle (nS) \rangle = 0$ and $\langle \Sigma_n \rangle =
\langle (n\Sigma) \rangle = 0$; hence it does not contribute: $ \langle
a_\ell \rangle= 0$.

\section{The 2PN spin-orbit equations of precession}\label{secVI}

In this Section we give the equations of evolution of the spins, or
precession equations, at relative 2PN order, \textit{i.e.} one PN order
beyond the dominant term. The precession equations are quite simple to
derive from the equation of parallel transport~\eqref{parallel}, which
we recall is valid at the \textit{linear} order in the spins [neglecting
$\mathcal{O}\left(S^2\right)$], but at \textit{any} PN order in that
term which is linear in the spins. The PN corrections are easily
computed from the non-spin part of the metric and Christoffel symbols
computed in Ref.~\cite{BFP98}. The precession equations in a general
frame take the form
\begin{equation}\label{preceq}
\frac{d \mathbf{S}_1}{d t}=\frac{1}{c^2}
\mathop{\mathbf{T}}_\text{S}{}_{\!\mathrm{1PN}}
+\frac{1}{c^3}\mathop{\mathbf{T}}_\text{SS}{}_{\!\mathrm{1.5PN}}
+\frac{1}{c^4}\mathop{\mathbf{T}}_\text{S}{}_{\!\mathrm{2PN}}
+\mathcal{O}\left(\frac{1}{c^5}\right)\,,
\end{equation}
together with the equation with $1\leftrightarrow 2$. At the lowest
order we find
\begin{equation}\label{T1PN}
\mathop{\mathbf{T}}_\text{S}{}_{\!\mathrm{1PN}} = \frac{G
  m_2}{r_{12}^2}\bigg[\mathbf{S}_1 (n_{12}v_{12}) - 2\mathbf{n}_{12}
  (v_{12}S_1) + (\mathbf{v}_1-2\mathbf{v}_2) (n_{12}S_1)\bigg] \, .
\end{equation}
The above equation is already known~\cite{K95, TOO01}. See \textit{e.g.}
Eq.~(4.3) in~\cite{TOO01} and the paragraph afterward commenting about
the difference with formulations based on an alternative definition for
the spin, like that of Ref.~\cite{K95}. The spin-spin (SS) term is also
known but is out of the scope of the present paper (and the parallel
transport equation we employ); it can be found elsewhere, see
Eqs.~(2)--(3) of~\cite{BCV03b}. Then we find that the next-order
spin-orbit term is
\begin{align}\label{T2PN}
  & \mathop{\mathbf{T}}_\text{S}{}_{\!\mathrm{2PN}} = \frac{G m_2}{r_{12}^2}
  \bigg\{ \mathbf{S}_1 \bigg[ (n_{12}v_2) (v_{12}v_2) - \frac{3}{2}
  (n_{12}v_2)^2 (n_{12} v_{12}) + \frac{G m_1}{r_{12}} (n_{12}v_1) \nonumber -
  \frac{G m_2}{r_{12}} (n_{12}v_{12}) \Big] \\ & + \mathbf{n}_{12} \bigg[
  (v_{12} S_1) \Big( 3 (n_{12}v_2)^2 + 2 (v_{12}v_2) \Big) + \frac{G
    m_1}{r_{12}} \Big(-16 (n_{12} S_1) (n_{12}v_{12}) + 3 (v_1 S_1) - 7 (v_2
  S_1)\Big) \nonumber \\ & \qquad + 2 (n_{12}S_1) \frac{G m_2}{r_{12}}
  (n_{12}v_{12}) \bigg] - \mathbf{v}_1 \bigg[\frac{3}{2} (n_{12}S_1)
  (n_{12}v_2)^2 + (v_{12} S_1) (n_{12}v_2) \nonumber \\ & \qquad - (n_{12}
  S_1) \frac{G}{r_{12}} (6 m_1 - m_2)\bigg] + \mathbf{v}_2 \bigg[ (n_{12}S_1)
  \Big( 2 (v_{12}v_2) + 3 (n_{12}v_2)^2 \Big) \nonumber \\ & \qquad + 2
  (n_{12}v_{12}) \Big((v_1 S_1) + (v_2 S_1)  \Big) - 5 (n_{12} S_1)
  \frac{G}{r_{12}} (m_1 - m_2 ) \bigg] \bigg\}\, .
\end{align}

For completeness and for the benefit of users of these formulas in the
data analysis of detectors, we present also the precession equations in
the CM frame, using our specific spin variables defined by
\eqref{SSigma}. These are
\begin{subequations}\label{preceqSSigma}\begin{align}
\frac{d \mathbf{S}}{d t} &= \frac{1}{c^2}
\mathop{\mathbf{U}}_\text{S}{}_{\!\mathrm{1PN}}
+\frac{1}{c^3}\mathop{\mathbf{U}}_\text{SS}{}_{\!\mathrm{1.5PN}}
+\frac{1}{c^4}\mathop{\mathbf{U}}_\text{S}{}_{\!\mathrm{2PN}}
+\mathcal{O}\left(\frac{1}{c^5}\right)\,,\\ \frac{d \mathbf{\Sigma}}{d
t} &= \frac{1}{c^2} \mathop{\mathbf{V}}_\text{S}{}_{\!\mathrm{1PN}}
+\frac{1}{c^3}\mathop{\mathbf{V}}_\text{SS}{}_{\!\mathrm{1.5PN}}
+\frac{1}{c^4}\mathop{\mathbf{V}}_\text{S}{}_{\!\mathrm{2PN}}
+\mathcal{O}\left(\frac{1}{c^5}\right)\,,
\end{align}\end{subequations}
where all the spin-orbit coefficients are given by
\begin{subequations}\label{UPN}\begin{align}
\mathop{\mathbf{U}}_\text{S}{}_{\!\mathrm{1PN}} &=
\frac{G\,m\,\nu}{r^2}\biggl\{\mathbf{n}\left[-4 (vS)-2\frac{\delta
m}{m}\,(v\Sigma)\right]\nonumber\\&\quad +\mathbf{v}\left[
3(nS)+\frac{\delta m}{m}\,(n\Sigma)\right]+(nv)\left[
2\mathbf{S}+\frac{\delta m}{m}\,\mathbf{\Sigma}\right]\biggr\}\,,\\
\mathop{\mathbf{U}}_\text{S}{}_{\!\mathrm{2PN}} &=
\frac{G\,m\,\nu}{r^2}\biggl\{\mathbf{n}\left[(vS)\left(-2v^2
+3(nv)^2-6\nu (nv)^2+7\frac{G m}{r}-8\nu\frac{G m}{r}\right)-14\frac{G
m}{r}(nS) (nv)\right.\nonumber\\&\quad\quad\left. + \frac{\delta
m}{m}\,(v\Sigma)\,\nu\left(-3(nv)^2-4\frac{G m}{r}\right)+ \frac{\delta
m}{m} \, \frac{G m}{r}
\,(n\Sigma)\,(nv)\left(2-\frac{\nu}{2}\right)\right]\nonumber\\&\quad 
+\mathbf{v}\left[(nS)\left(2v^2-4\nu v^2-3(nv)^2+\frac{15}{2}\nu
(nv)^2+4\frac{G m}{r}-6\nu\frac{G
m}{r}\right)+(vS)(nv)\left(2-6\nu\right)\right.\nonumber\\&\quad\quad\left.
+ \frac{\delta m}{m}\,(n\Sigma)\,\left(-\frac{3}{2} \nu v^2+3\nu
(nv)^2-\frac{G m}{r}-\frac{7}{2}\nu\frac{G m}{r}\right) -3\frac{\delta
m}{m}\,(v\Sigma)\,(nv)\,\nu\right]\nonumber\\&\quad
+\mathbf{S}\,(nv)\left[v^2-2\nu v^2-\frac{3}{2}(nv)^2+3\nu
(nv)^2-\frac{G m}{r}+2\nu\,\frac{G m}{r}\right]\nonumber\\&\quad
+\frac{\delta m}{m}\,\mathbf{\Sigma}\,(nv)\left[-\nu
v^2+\frac{3}{2}\nu\,(nv)^2-\frac{G m}{r}+\nu\,\frac{G m}{r}\right]\biggr\}\,,
\end{align}\end{subequations}
and
\begin{subequations}\label{VPN}\begin{align}
\mathop{\mathbf{V}}_\text{S}{}_{\!\mathrm{1PN}} &=
\frac{G\,m}{r^2}\biggl\{\mathbf{n}\left[(v\Sigma)\left(-2+4\nu\right)-
2\frac{\delta m}{m}\,(vS)\right]\nonumber\\&\quad +\mathbf{v}\left[
(n\Sigma)\left(1-\nu\right)+\frac{\delta m}{m}\,(nS)\right]+(nv)\left[
\mathbf{\Sigma}\left(1-2\nu\right)+\frac{\delta
m}{m}\,\mathbf{S}\right]\biggr\}\,,\\
\mathop{\mathbf{V}}_\text{S}{}_{\!\mathrm{2PN}} &=
\frac{G\,m}{r^2}\biggl\{\mathbf{n}\left[(v\Sigma)\,\nu\left(-2v^2 +6\nu
(nv)^2+3\frac{G m}{r}+8\nu\frac{G
m}{r}\right)\right.\nonumber\\&\quad\quad\left. +\frac{G
m}{r}(n\Sigma)\,(nv)\left(2-\frac{45}{2}\nu+2\nu^2\right)
\right.\nonumber\\&\quad\quad\left. + \frac{\delta
m}{m}\,(vS)\,\nu\left(-3(nv)^2-4\frac{G m}{r}\right)+ \frac{\delta
m}{m}\,\frac{G
m}{r}\,(nS)\,(nv)\left(2-\frac{\nu}{2}\right)\right]\nonumber\\&\quad
+\mathbf{v}\left[(n\Sigma)\left(\frac{\nu}{2}v^2 + 2\nu^2
v^2-\frac{9}{2}\nu^2 (nv)^2 - \frac{G m}{r}+ \frac{9}{2}\nu\frac{G m}{r}
+ 8 \nu^2\frac{G m}{r}\right)\right.\nonumber\\&\quad\quad\left.
+(v\Sigma)(nv)\nu\left(-1+6\nu\right)-3\frac{\delta
m}{m}\,(vS)\,(nv)\,\nu\right.\nonumber\\&\quad\quad\left. +
\frac{\delta m}{m}\,(nS)\,\left(-\frac{3}{2} \nu v^2+3\nu (nv)^2-\frac{G
m}{r}-\frac{7}{2}\nu\frac{G m}{r}\right) \right]\nonumber\\&\quad
+\mathbf{\Sigma}\,(nv)\left[2\nu^2 v^2-3\nu^2 (nv)^2-\frac{G
m}{r}+4\nu\,\frac{G m}{r}-2\nu^2\,\frac{G m}{r}\right]\nonumber\\&\quad
+\frac{\delta m}{m}\,\mathbf{S}\,(nv)\left[-\nu
v^2+\frac{3}{2}\nu\,(nv)^2-\frac{G m}{r}+\nu\,\frac{G
m}{r}\right]\biggr\}\,.
\end{align}\end{subequations}
To these expressions one may add the SS terms in the standard way (see
Eqs.~(2)--(3) of~\cite{BCV03b}).

\section{Spin effects in the conserved integrals of the motion}\label{secVII}

Having obtained in Sec.~\ref{secV} the equations of motion, the
important task is now to deduce from them the complete set of conserved
integrals of the motion associated with the global Poincar\'e invariance
of these equations (which has been checked in Ref.~\cite{TOO01} and
Appendix~\ref{secA} below). In principle, the conserved integrals of the
motion, which generalize the usual notions of energy, angular and linear
momenta, and center of mass position, should be best derived from a
Lagrangian. In the present paper, however, we did not attempt to derive
a complete Lagrangian for the particles with spins (see~\cite{BI80} for
a discussion on how to formulate Lagrangians with spins); rather, we
have obtained the integrals of the motion by ``guess-work'', starting
from their most general admissible form, and then imposing the
conservation laws when the equations of motion are
satisfied.\,\footnote{As usual we neglect the radiation reaction effect
at 2.5PN order. Indeed we know that such an effect does not depend on
the spins. The contribution of the spins in the radiation reaction force
comes in at 1.5PN order beyond the dominant effect, which means at the
4PN level, and has been computed in Ref.~\cite{W05}. Radiation reaction
effects will be included into the present formalism when we obtain the
contributions of the spins in the GW flux \cite{BBF06spin}.} Here we
simply state the results.

The PN expansion of the conserved integral of the energy, namely $E$
such that $d E/dt=0$, reads as
\begin{equation}\label{E}
E = E_\mathrm{N}+\frac{1}{c^2}E_\mathrm{1PN}+\frac{1}{c^3}
\mathop{E}_\text{S}{}_{\!\mathrm{1.5PN}}
+\frac{1}{c^4}\left[E_\mathrm{2PN}+
\mathop{E}_\text{SS}{}_{\!\mathrm{2PN}}\right]
+\frac{1}{c^5}\mathop{E}_\text{S}{}_{\!\mathrm{2.5PN}}
+\mathcal{O}\left(\frac{1}{c^6}\right) \, ,
\end{equation}
where the non-spin pieces, $E_\mathrm{N}$, $E_\mathrm{1PN}$ and
$E_\mathrm{2PN}$, are known and can be found \textit{e.g.} in
Ref.~\cite{ABF01}. For instance we have
$E_\mathrm{N}=\frac{1}{2}m_1\,v_1^2 + \frac{1}{2}m_2\,v_2^2-
\frac{G\,m_1\,m_2}{r_{12}}$. For the lowest-order spin-orbit effect we find, in
agreement with the standard result,
\begin{subequations}\label{ES}\begin{equation}\label{E15PN}
\mathop{E}_\text{S}{}_{\!\mathrm{1.5PN}} = \frac{G
m_2}{r_{12}^2}\,(S_1,n_{12},v_1) + 1\leftrightarrow 2 \, ,
\end{equation}
where we employ a special notation for the totally anti-symmetric
``mixed product'' between three vectors, as given
in~\eqref{eq:dependence}. For the spin-orbit contribution at 2.5PN order
we find
\begin{align}\label{E25PN}
\mathop{E}_\text{S}{}_{\!\mathrm{2.5PN}} = \frac{G m_2}{r_{12}^2} &
  \bigg[(S_1,n_{12},v_1)\bigg(\frac{1}{2}v_1^2 -3v_2^2
  +3(n_{12}v_1)(n_{12}v_2)+\frac{3}{2}(n_{12}v_2)^2-2\frac{G\,m_1}{r_{12}}
  +\frac{G\,m_2}{r_{12}}\bigg) \nonumber\\& + (S_1,n_{12},v_2)\bigg(2
  v_1^2-(v_1v_2)+2v_2^2 -3(n_{12}v_1)^2-3(n_{12}v_1)(n_{12}v_2)-3\frac{G
  m_2}{r_{12}}\bigg)\nonumber\\& + (n_{12},v_1,v_2)\Bigl(
  (v_1S_1)+2(v_2S_1)\Bigr)\bigg] + 1\leftrightarrow 2 \, .
\end{align}
\end{subequations}
Notice that several equivalent forms can be given to this result. For
instance if wished one could introduce the mixed product $(S_1,v_1,v_2)$
in place of a $(n_{12},v_1,v_2)$ in the last term of~\eqref{E25PN},
making use of linear combinations such as
$(n_{12}v_1)(S_1,v_1,v_2)=(n_{12},
v_1,v_2)(v_1S_1)+(S_1,n_{12},v_2)v_1^2 -(S_1,n_{12},v_1)(v_1v_2)$ [a
consequence of Eq.~\eqref{eq:dependence}]. As before we do not give the
SS contribution at 2PN order (see~\cite{K95} for instance).

We give here the corresponding result for the conserved
center-of-mass energy in the CM frame:
\begin{equation}\label{enstruct}
E = m\,\nu\,c^2\,\left\{e_\mathrm{N}+\frac{1}{c^2}e_\mathrm{1PN}
+\frac{1}{c^3}\mathop{e}_{\text{S}}{}_{\!\mathrm{1.5PN}}
+\frac{1}{c^4}\left[e_\mathrm{2PN}+
\mathop{e}_{\text{SS}}{}_{\!\mathrm{2PN}}\right]  
+\frac{1}{c^5}\mathop{e}_{\text{S}}{}_{\!\mathrm{2.5PN}} +
\mathcal{O}\left(\frac{1}{c^6}\right)\right\}\,,
\end{equation}
where $e_\mathrm{N}=\frac{1}{2}v^2-\frac{G m}{r}$
(see~\cite{BI03CM} for the other non-spin contributions). The SO
coupling terms in the CM frame are found to be
\begin{subequations}\label{en25PN}\begin{align}
\mathop{e}_{\text{S}}{}_{\!\mathrm{1.5PN}}&=\frac{G}{r^2}\left\{(S,n,v)
+ (\Sigma,n,v)\,\frac{\delta m}{m}\right\}\,,\\
\mathop{e}_{\text{S}}{}_{\!\mathrm{2.5PN}}&=\frac{G}{r^2}
\left\{(S,n,v)\left[-\frac{3}{2}(1+\nu)v^2 -\frac{3}{2}\nu
(nv)^2+\frac{G\,m}{r}\right]\right.\nonumber\\ &\qquad\left. +
(\Sigma,n,v)\left[\frac{1}{2}(1-5\nu)v^2 +
\frac{1}{2}(2+\nu)\frac{G\,m}{r}\right]\frac{\delta m}{m}\right\}\,.
\end{align}\end{subequations}

Let us next deal with the conserved total angular momentum $\mathbf{J}$,
\textit{i.e.} $d\mathbf{J}/dt=\mathbf{0}$, sum of orbital and spin
contributions, which we write as
\begin{equation}\label{J}
\mathbf{J} = \mathbf{L} +\frac{1}{c}\mathbf{S}_1
+\frac{1}{c}\mathbf{S}_2 \, ,
\end{equation}
where $\mathbf{L}$ is the orbital angular momentum, and where
$\mathbf{S}_1$ and $\mathbf{S}_2$ are the contravariant spin vectors
defined following the specific choice made in Eq.~\eqref{SAcov} [recall
also their peculiar dimension which follows from~\eqref{S}]. The angular
momentum $\mathbf{L}$ admits the PN expansion
\begin{equation}\label{L}
\mathbf{L} = \mathbf{L}_\mathrm{N}+\frac{1}{c^2}\mathbf{L}_\mathrm{1PN}+
\frac{1}{c^3} \mathop{\mathbf{L}}_\text{S}{}_{\!\mathrm{1.5PN}}
+\frac{1}{c^4}\left[\mathbf{L}_\mathrm{2PN}+
\mathop{\mathbf{L}}_\text{SS}{}_{\!\mathrm{2PN}}\right]
+\frac{1}{c^5}\mathop{\mathbf{L}}_\text{S}{}_{\!\mathrm{2.5PN}}
+\mathcal{O}\left(\frac{1}{c^6}\right) \, ,
\end{equation}
where all the non-spin pieces are given by Eq.~(4.4) of~\cite{ABF01}.
For instance, $\mathbf{L}_\mathrm{N}=m_1\,\mathbf{y}_1\times
\mathbf{v}_1 + m_2\,\mathbf{y}_2\times \mathbf{v}_2$. Now, in order to
express in the best way the spin-orbit contributions in $\mathbf{L}$, we find
that they must be written in the following way,
\begin{subequations}\label{Ldecomp}\begin{align}
\mathop{\mathbf{L}}_\text{S}{}_{\!\mathrm{1.5PN}} &= \mathbf{y}_1\times
\mathop{\mathbf{p}}_\text{S}{}_{\!1} + \mathbf{y}_2\times
\mathop{\mathbf{p}}_\text{S}{}_{\!2} +
\mathop{\mathbf{K}}_\text{S}{}_{\!\mathrm{1.5PN}} \, ,\label{Ldecompa}\\
\mathop{\mathbf{L}}_\text{S}{}_{\!\mathrm{2.5PN}} &= \mathbf{y}_1\times
\mathop{\mathbf{q}}_\text{S}{}_{\!1} + \mathbf{y}_2\times
\mathop{\mathbf{q}}_\text{S}{}_{\!2} +
\mathop{\mathbf{K}}_\text{S}{}_{\!\mathrm{2.5PN}}\,,
\end{align}\end{subequations}
in which we have introduced some convenient notions of the ``individual
linear momenta'' of the particles, say ${}_\text{S}\mathbf{p}_1$ and
${}_\text{S}\mathbf{p}_2$ at 1.5PN order, and ${}_\text{S}\mathbf{q}_1$
and ${}_\text{S}\mathbf{q}_2$ at 2.5PN order. The extra terms in the
RHS, ${}_\text{S}\mathbf{K}_{\mathrm{1.5PN}}$ and
${}_\text{S}\mathbf{K}_{\mathrm{2.5PN}}$, incorporate all what remains,
the point being that they depend on the positions of the particles only
through their \textit{relative} separation, \textit{i.e.}
$r_{12}=\vert\mathbf{y}_1-\mathbf{y}_2\vert$ and
$\mathbf{n}_{12}=(\mathbf{y}_1-\mathbf{y}_2)/r_{12}$. The only
dependence of the conserved angular momentum on the individual positions
$\mathbf{y}_1$ and $\mathbf{y}_2$ is the one which is given explicitly
by the first terms of Eqs.~\eqref{Ldecomp}.\,\footnote{However, let us
stress that the definition of some individual momenta for the particles
is merely introduced here as a convenient notation. In order to define
in a meaningful way the notions of individual linear momenta of the
particles with spins, we would need a Lagrangian, which as said before
we did not compute, and the linear momenta would simply be the conjugate
momenta of the ordinary positions.} The results we find for these
momenta are
\begin{subequations}\label{pq}\begin{align}
\mathop{\mathbf{p}}_\text{S}{}_{\!1} &=
-\frac{G\,m_2}{r_{12}^2}\,\mathbf{n}_{12}\times \mathbf{S}_1\,,\\
\mathop{\mathbf{q}}_\text{S}{}_{\!1} &= \frac{G m_2}{r_{12}^2} \bigg\{
\mathbf{n}_{12} \times \mathbf{S}_1 \bigg[- \frac{5}{2} v_1^2 + 4
  (v_1v_2) - 2 v_2^2 + \frac{3}{2} (n_{12}v_2)^2 + \frac{2 G}{r_{12}}
  (m_1+m_2) \bigg] \nonumber \\ & \qquad + 3 \mathbf{v}_{12} \times
\mathbf{S}_1 (n_{12} v_1) + \mathbf{n}_{12} \times \mathbf{v}_1 (v_1
S_1) \nonumber \\ & \qquad + \mathbf{n}_{12} \times \mathbf{v}_{12}
\bigg[ 3 (n_{12} S_1) \Big( (n_{12} v_1) + (n_{12} v_2) \Big) + (v_{12}
  S_1) \bigg] \bigg\} \,,
\end{align}\end{subequations}
together with the equations with $1\leftrightarrow 2$. The last terms in
the RHS of Eqs.~\eqref{Ldecomp} are explicitly given by
\begin{subequations}\label{K}\begin{align}
\mathop{\mathbf{K}}_\text{S}{}_{\!\mathrm{1.5PN}} &=
  \frac{G\,m_2}{r_{12}}\biggl[2(n_{12}S_1)\,\mathbf{n}_{12}-\mathbf{S}_1\biggr]
  + \frac{1}{2}v_1^2 \mathbf{S}_1-(v_1S_1)\mathbf{v}_1 +
  1\leftrightarrow 2 \, ,\\
  \mathop{\mathbf{K}}_\text{S}{}_{\!\mathrm{2.5PN}} &= \mathbf{S}_1
  \bigg[ \frac{3}{8} v_1^4 + \frac{G m_2}{r_{12}} \bigg(-\frac{1}{2}
  v_1^2 + 3 (v_{12}v_2) + 3 (n_{12}v_1)^2 - 4 (n_{12} v_1) (n_{12} v_2)
  \nonumber \\ & \qquad + \frac{3}{2} (n_{12}v_2)^2 + \frac{1}{2}
  \frac{G m_1}{r_{12}} + \frac{3}{2} \frac{G m_2}{r_{12}} \bigg) \bigg]
  + \frac{G m_2 \mathbf{n}_{12}}{r_{12}} \bigg[ (n_{12} S_1) \bigg( 3
  v_{12}^2 + (v_1v_2) \nonumber \\ & \qquad - 3 (n_{12} v_1)^2 - 4
  \frac{G m_1}{r_{12}} - \frac{G m_2}{r_{12}} \bigg) + (v_{12}S_1)\Big(
  - 3 (n_{12} v_1) + (n_{12} v_2) \Big)\bigg] \nonumber \\ & +
  \mathbf{v}_1 \bigg[- \frac{1}{2} (v_1S_1) v_1^2 + \frac{G m_2}{r_{12}}
  \bigg(3 (n_{12} S_1) (n_{12} v_2) - 2 (v_1 S_1) \bigg) \bigg]
  \nonumber \\ & + \frac{G m_2 \mathbf{v}_2}{r_{12}} \bigg[- 3
  (n_{12}S_1) (n_{12} v_2) + 3 (v_1 S_1) + 4 (v_2 S_1) \bigg] + 1
  \leftrightarrow 2 \, .
\end{align}\end{subequations}
The 1.5PN term in the conserved angular momentum, Eq.~\eqref{Ldecompa},
agrees with the result of Kidder~\cite{K95}.\,\footnote{Ref.~\cite{K95}
uses different definitions for the spin variables, which are related to
ours by
$$\left(\mathbf{S}_1\right)_\text{Kidder}=\left(1+\frac{G
m_2}{c^2r_{12}}\right)\mathbf{S}_1-\frac{1}{2c^2}(v_{1}S_1)\,\mathbf{v}_{1}\,.$$}

Let us add a comment on the meaning of the conservation of the total
angular momentum $\mathbf{J}$ at 2.5PN order [Eq.~\eqref{J} with
\eqref{L}]. When differentiating $\mathbf{J}$ with respect to time, we
generate several spin contributions at 2.5PN order: (i) The ``main'' one
is coming from the differentiation of the Newtonian term
$\mathbf{L}_\mathrm{N}$, and is due to the replacement of the
acceleration by the equations of motion~\eqref{a1struct} with
\eqref{A2.5PN}; (ii) There is the one coming from the differentiation of
the 1PN part $\mathbf{L}_\mathrm{1PN}$, since the replacement of the
accelerations at order 1.5PN [Eq.~\eqref{A1.5PN}] therein does also
produce some terms at 2.5PN order; (iii) When differentiating the
lowest-order spin-orbit term ${}_\text{S}\mathbf{L}_\mathrm{1PN}$, the
derivative of the spins gives other 2.5PN terms \textit{via} the
precessional equations; (iv) When differentiating the spin vectors
themselves, $\mathbf{S}_1$ and $\mathbf{S}_2$, one must make use of the
precessional equations with their full 2PN accuracy\,\footnote{This is
the only place where one needs the precessional equations with 2PN
accuracy.} which are given by Eqs.~\eqref{preceq}--\eqref{T2PN}. Only
when account is taken of all these replacements (i)--(iv) of
accelerations and spin precession, does one find that $\mathbf{J}$
\textit{is} conserved, $d \mathbf{J}/d t = \mathbf{0}$, up to 2.5PN
order (neglecting the 2.5PN non-spin radiation reaction damping).

The orbital angular momentum in the CM frame reads:
\begin{equation}\label{lstruct}
\mathbf{L} = \nu \left\{\bm{\ell}_{\!\mathrm{N}}+ \frac{1}{c^2}
\bm{\ell}_{\!\mathrm{1PN}}+ \frac{1}{c^3}
\mathop{\bm{\ell}}_{\text{S}}{}_{\!\mathrm{1.5PN}} + \frac{1}{c^4}\left[
\bm{\ell}_{\!\mathrm{2PN}} +
\mathop{\bm{\ell}}_{\text{SS}}{}_{\!\mathrm{2PN}}\right] + \frac{1}{c^5}
\mathop{\bm{\ell}}_{\text{S}}{}_{\!\mathrm{2.5PN}} +
\mathcal{O}\left(\frac{1}{c^6}\right)\right\}\,,
\end{equation}
where $\bm{\ell}_{\!\mathrm{N}}= m \,\mathbf{x} \times \mathbf{v}$; the
non-spin contributions can be found in Refs.~\cite{DD81b, ABF01,
BI03CM}. We have
\begin{subequations}\label{l15PN}\begin{align}
\mathop{\bm{\ell}}_{\text{S}}{}_{\!\mathrm{1.5PN}} &=
\frac{1}{2}\frac{\delta m}{m}\,v^2\,\mathbf{\Sigma} + \left (
\frac{v^2}{2} - \frac{G\,m}{r} \right )\,\mathbf{S} + \left [ 3
\frac{G\,m}{r^3}\,(x\, S) + \frac{G\,m}{r^3}\,(x \,\Sigma)\,
\frac{\delta m}{m} \right ] \mathbf{x} \nonumber \\ &+ \left [ - (v\,S)
- (v\,\Sigma)\,\frac{\delta m}{m} \right ]\,\mathbf{v}\,,\\
\mathop{\bm{\ell}}_{\text{S}}{}_{\!\mathrm{2.5PN}} &= -
\frac{G\,m}{r^3}\,\frac{\delta m}{m}\,\nu\,(\Sigma,x,v)\, \mathbf{x}
\times \mathbf{v} \nonumber \\ & + \; \mathbf{\Sigma}\,\left [ \left
(-\frac{1}{2} + \frac{\nu}{2} \right )\, \frac{G^2\,m^2}{r^2} + \left (
2 - \frac{\nu}{2} \right)\,v^2\,\frac{G\,m}{r} + \left ( \frac{3}{8} -
\frac{5}{4}\,\nu \right )\,v^4 + 3\nu\, \frac{G\,m}{r^3}\,(x\,v)^2
\right ]\,\frac{\delta m}{m} \nonumber \\ & + \; \mathbf{S}\, \left [
\left ( \frac{3}{2} + \frac{7}{2}\,\nu \right )\,\frac{G\,m}{r^3}\,(x
\,v)^2 + \left ( -1 - \frac{\nu}{2} \right )\,\frac{G\,m}{r}\,v^2 +
\left ( \frac{3}{8} - \frac{9}{8}\,\nu \right )\,v^4\right ] \nonumber
\\ & + \; \mathbf{x}\,\left [ \left ( \frac{1}{2} + \frac{\nu}{2} \right
)\,\frac{G\,m}{r^3}\,v^2\,(x\,\Sigma) + \left ( 1 - \frac{\nu}{2} \right
)\,\frac{G^2\,m^2}{r^4}\,(x\,\Sigma) - 3 \nu\,\frac{G\,m}{r^4}\,
(x\,v)^2\,(x\,\Sigma) \right. \nonumber \\ & \qquad + \left. \left ( -1
- \frac{5\nu}{2} \right )\,\frac{G\,m}{r^3}\,(x\,v)\,(v\,\Sigma) \right
]\,\frac{\delta m}{m} \nonumber \\ & + \; \mathbf{x}\,\left [
-3\,\frac{G^2\,m^2}{r^4}\,(x\,S) + \left ( \frac{7}{2} - \frac{\nu}{2}
\right )\,\frac{G\,m}{r^3}\,v^2\,(x\,S) -
\frac{9\nu}{2}\,\frac{G\,m}{r^4}\,(x\,v)^2\,(x\,S) \right. \nonumber \\
& \left. \qquad + \; \left ( -2 - 3\nu \right )\,
\frac{G\,m}{r^3}\,(x\,v)\,(v\,S) \right ] \nonumber \\ & +
\mathbf{v}\,\left [ -4\, \frac{G\,m}{r}\,\nu\,(v\,\Sigma) + \left (
-\frac{1}{2} + 2 \nu \right)\,v^2\,(v\,\Sigma) -
4\nu\,\frac{G\,m}{r}\,(v\,\Sigma) + \frac{5
\nu}{2}\frac{G\,m}{r^3}\,(x\,\Sigma)\,(x\,v) \right ]\,\frac{\delta
m}{m} \nonumber \\ & + \mathbf{v}\,\left [ \left (- \frac{1}{2} +
\frac{3 \nu}{2} \right )\,v^2\,(v\,S) - 7\,\nu\,\frac{G\,m}{r}\,(v\,S) +
\left (6\,\nu - 3 \right )\,\frac{G\,m}{r^3}\,(x\,S)\,(x\,v) \right ]\,.
\end{align}\end{subequations}

Finally let us give the conserved integrals of the linear momentum
$\mathbf{P}$ and center of mass position $\mathbf{G}$, which are related
to each other by $d \mathbf{G}/d t = \mathbf{P}$. Recall that the
existence of the center-of-mass integral $\mathbf{G}$ is a consequence
of the boost-invariance of the equations of motion (\textit{cf.}
Appendix~\ref{secA}). Both $\mathbf{P}$ and $\mathbf{G}$ admit a PN
expansion exactly like those of $E$ and $\mathbf{L}$. Quite naturally,
we find that the spin-orbit contributions in $\mathbf{P}$ are simply
given by the sum of the ``individual'' linear momenta for each particles
that we found convenient to introduce in order to express the angular
momentum in Eqs.~\eqref{Ldecomp}. Thus,
\begin{subequations}\label{P}
\begin{align}
\mathop{\mathbf{P}}_\text{S}{}_{\!\mathrm{1.5PN}} &=
\mathop{\mathbf{p}}_\text{S}{}_{\!1} +
\mathop{\mathbf{p}}_\text{S}{}_{\!2} \, ,\\
\mathop{\mathbf{P}}_\text{S}{}_{\!\mathrm{2.5PN}} &=
\mathop{\mathbf{q}}_\text{S}{}_{\!1} +
\mathop{\mathbf{q}}_\text{S}{}_{\!2} \, ,
\end{align}
\end{subequations}
where the explicit expressions~\eqref{pq} hold. For $\mathbf{G}$, we
obtain rather simple expressions,
\begin{subequations}\label{G}\begin{align}
\mathop{\mathbf{G}}_\text{S}{}_{\!\mathrm{1.5PN}} &= \mathbf{v}_1 \times
\mathbf{S}_1 + \mathbf{v}_2 \times \mathbf{S}_2 \, , \\
\mathop{\mathbf{G}}_\text{S}{}_{\!\mathrm{2.5PN}} &= \frac{1}{2}
\mathbf{v}_1 \times \mathbf{S}_1 v_1^2 - \frac{G m_2}{r_{12}} \bigg\{
-\frac{\mathbf{y}_1}{r_{12}} (S_1, n_{12}, v_1) - 2 \mathbf{v}_1 \times
\mathbf{S}_1 + 3 \mathbf{v}_2 \times \mathbf{S}_1 \nonumber \\ &
\qquad\quad + \Big(\mathbf{n}_{12} \times \mathbf{v}_1 + \mathbf{n}_{12}
\times \mathbf{v}_2\Big) (n_{12} S_1) \bigg\} + 1 \leftrightarrow 2 \, .
\end{align}
\end{subequations}

The derivation of the complete set of integrals of the motion gives us
further confidence in the physical soundness of the equations of motion
derived in this paper. Those results, together with the analyses
performed in Appendices ~\ref{secA} and~\ref{secB}, complete the
resolution of the problem of linear spin-orbit effects in the binary's
equations of motion at 2.5PN order.


\appendix

\section{Lorentz invariance of the equations of motion}\label{secA}

Because of the global Poincar\'e invariance of the Einstein equations
(with bounded sources), and the manifest covariance of the De Donder
harmonicity condition, it is not possible to physically distinguish
between two harmonic-coordinate grids differing by a mere Lorentz
transformation. As a result, the equations of motion must be of the same
form in two such grids. In other words, up to an arbitrary PN order $ n
$, the link between the boosted acceleration $ \mathbf{a}'_1(
\mathbf{y}_C, \mathbf{v}_C, \mathbf{a}_C ) $ and the boosted positions $
\mathbf{y}'_B(\mathbf{y}_C, \mathbf{v}_C) $, velocities, $
\mathbf{v}'_B( \mathbf{y}_C, \mathbf{v}_C) $ and spins $ \mathbf{S}'_1(
\mathbf{y}_C, \mathbf{v}_C, \mathbf{S}_C) $, must be given by the
original equations of motion [\emph{i.e.} Eq.~\eqref{a1struct} at the
2.5PN level] with the original variables being replaced by their primed
counterparts. Note that the Euclidean metric and the totally
antisymmetric tensors remain unchanged under Lorentz transformations.
Schematically, we may write $\mathbf{a}'_1 = \mathbf{A}(\mathbf{y}'_B,
\mathbf{v}'_B, \mathbf{S}'_B, \delta_{ij}, \varepsilon_{ijk}) $ for $ B
= 1, 2$. The resulting relation between un-boosted quantities,
\begin{equation}
\mathbf{a}'_1( \mathbf{y}_C, \mathbf{v}_C, \mathbf{a}_C ) = \mathbf{A}
\Big( \mathbf{y}'_B(\mathbf{y}_C, \mathbf{v}_C), \mathbf{v}'_B (
\mathbf{y}_C, \mathbf{v}_C), \mathbf{S}'_B ( \mathbf{y}_C, \mathbf{v}_C,
\mathbf{S}_C), \delta_{ij}, \varepsilon_{ijk} \Big) +
\mathcal{O}\Big(\frac{1}{c^{n+1}} \Big) \, ,
\end{equation}
defines a function $\mathbf{A}'$ as $ \mathbf{a}_1 =
\mathbf{A}'(\mathbf{y}_C, \mathbf{v}_C, \mathbf{S}_C, \delta_{ij},
\varepsilon_{ijk}) + \mathcal{O}( 1/c^{n+1} ) $. Equivalence with the
equations of motion in the un-boosted frame: $ \mathbf{a}_1 =
\mathbf{A}(\mathbf{y}_B, \mathbf{v}_B, \mathbf{S}_B, \delta_{ij},
\varepsilon_{ijk}) + \mathcal{O}(1/c^{n+1}) $, means precisely that
\begin{equation} \label{eq:Lorentz_invariance}
\mathbf{A} = \mathbf{A}'\, ,
\end{equation}
up to negligible PN corrections. This property constitutes
the so-called explicit Lorentz boost invariance of the equations of
motion. It happens to be a very powerful check for the coefficients
entering the functions $ \mathbf{A} $ of Eq.~\eqref{a1struct}, and in
particular its contribution $ \mathbf{A}_{2.5\text{PN}} $ [see
Eq.~\eqref{A2.5PN}].

In order to verify the validity of Eq.~\eqref{eq:Lorentz_invariance}, we
need to determine the function $ \mathbf{A}' $ explicitly, which
requires to know how $ \mathbf{y}_B $, $\mathbf{v}_B $, $\mathbf{a}_1$
and $ \mathbf{S}_B$ transform under a Lorentz boost. Let us start with
considering an arbitrary space-time event $P$ with coordinates $ x^\mu $
in the current working frame $(\mathcal{F})$. Its coordinates in a
boosted frame $ (\mathcal{F}') $ of relative velocity $ \mathbf{V} $ are
related to the original ones by $ x'^\mu = \Lambda^{\!\mu}_{\,
\nu}(\mathbf{V}) x^\nu $, where the Lorentz matrix $ \Lambda^{\!\mu}_{\,
\nu}(\mathbf{V}) $ is given by
\begin{subequations}
\begin{align}
& \Lambda^{\!0}_{\,0}(\mathbf{V}) = \gamma \, ,\\ &
\Lambda^{\!i}_{\,0}(\mathbf{V}) = \Lambda^{\!0}_{\,i}(\mathbf{V}) = -
\gamma \frac{V^i}{c} \, ,\\ & \Lambda^{\!i}_{\,j}(\mathbf{V}) =
\delta^i_j + \frac{\gamma^2}{\gamma + 1} \frac{V^i V_j}{c^2} \, ,
\end{align}
\end{subequations}
with $ \gamma $ being the Lorentz factor $ 1/\sqrt{1 - V^2/c^2}$. An
event $ Q $ with coordinates $ y'^\mu $ in $ (\mathcal{F}') $ is
simultaneous to $ P $ in the new frame if and only if $ y'^0 = x'^0 $.
There exist two such events located on the two world-lines of the binary
companions. Their coordinates in $ (\mathcal{F}') $ are denoted by $
{y'}_1^\mu = (c t', \mathbf{y}'_1) $ and $ {y'}_2^\mu = (c t',
\mathbf{y}'_2) $ respectively. The mapping $ t'\rightarrow \mathbf{y}'_1
$ defines a function $ \mathbf{y}'_1(t') $, and similarly for the second
body. The events having coordinates $(ct', \mathbf{y}'_1(t'))$ and
$(ct', \mathbf{y}'_2(t'))$ in $ (\mathcal{F}') $ do not generally appear
as simultaneous in $ (\mathcal{F}) $. They may be referred to in
components as $ (ct_1, \mathbf{y}_1(t_1)) $ and $ (ct_2,
\mathbf{y}_2(t_2)) $ in that frame, the functions $ \mathbf{y}_1(t) $
and $ \mathbf{y}_2(t) $ being the original trajectories. By
construction, we have
\begin{equation}
y'^\mu_1(t') = \Lambda^{\!\mu}_{\, \nu}(\mathbf{V}) y_1^\nu(t_1) \, .
\end{equation}
Let us express in the end the RHS in terms of the coordinate time $t$. A
derivation of the general formula linking $ \mathbf{y}'_1(t') $ to $
\mathbf{y}_1(t) $ in the PN scheme can be found in~\cite{BF01a}. This
relation reads, see Eqs.~(3.20) in~\cite{BF01a},
\begin{equation}
  \mathbf{y}'_1(t') = \mathbf{y}_1(t) - \gamma \mathbf{V} \Big(t -
  \frac{1}{c^2} \frac{\gamma}{\gamma + 1} (Vx)\Big) +
  \sum_{n=1}^{+\infty} \frac{(-1)^n}{c^{2n} n!} \partial_t^{n-1} \bigg[
  (Vr_1)^n \Big(\mathbf{v}_1 - \frac{\gamma}{\gamma + 1} \mathbf{V}
  \Big) \bigg] \, .
\end{equation}
The velocity and acceleration follow from the partial derivation with
respect to $ t' $ together with the formula $ \partial'_t = \gamma
\partial_t + \gamma V^i \partial_i $:
\begin{subequations}
\begin{align}
& \mathbf{v}'_1 = \frac{\mathbf{v}_1}{\gamma} - \mathbf{V} +
\frac{1}{\gamma} \sum_{n=1}^{+\infty} \frac{(-1)^n}{c^{2n} n!}
\partial_t^n \bigg[(Vr_1)^n \Big(\mathbf{v}_1 - \frac{\gamma}{\gamma +
1} \mathbf{V}\Big) \bigg] \, , \\ & \mathbf{a}'_1 = \frac{1}{\gamma^2}
\bigg\{ \mathbf{a}_1 + \sum_{n=1}^{+\infty} \frac{(-1)^{n}}{c^{2n} n!}
\partial_t^{n+1} \bigg[(Vr_1)^n \Big(\mathbf{v}_1 - \frac{\gamma}{\gamma
+ 1} \mathbf{V}\Big) \bigg] \bigg\} \, .
\end{align}
\end{subequations}

The spin components in the new frame cannot be obtained directly from
the linear Lorentz transformation law. This is because the definition of
$\mathbf{S}_1$ and $\mathbf{S}_2$ involves the inverse of the 3-metric $
\gamma_{ij} $ induced by $ g_{\mu\nu} $ on a slice $ t = \text{const.}$
Now, $ \gamma_{ij} $ implicitly depends on the choice of the coordinate
time and is generally singular because of the particle's
self-gravitation. To avoid complications rising from this second issue,
we shall first focus on the case of test particles on a fixed
background.

In the frame $ (\mathcal{F}') $, the spin components of the first test body
read
\begin{equation}
S_1'^i(t') = \gamma'^{ij}_1(t') S'^1_j(t')
\end{equation}
with $ \gamma'^{ij}_1(t') = \gamma'^{ij}(\mathbf{y}'_1, t') $. Whereas
the transformation law of $ \gamma^{ij}_1 $ is more difficult, that of $
(\gamma_{ij})_1 = (g_{ij})_1 $ results straightforwardly from the
transformation of the space-time metric:
\begin{equation} \label{eq:primed_metric}
(g'_{ij})_1(t') = \Lambda_{i}^{\,\,\mu}(\mathbf{V})
\Lambda_{j}^{\,\,\nu}(\mathbf{V}) (g_{\mu\nu})_1(t_1) \, ,
\end{equation}
with $\Lambda_\lambda^{\,\,\mu}(\mathbf{V}) = \Lambda^{\!\mu}_{\,
\lambda}(-\mathbf{V}) $ denoting the inverse transformation. Therefore,
computing $ \gamma'^{ij}_1(t') $ amounts to expressing the latter
quantity as a function of $ (g'_{ij})_1 $. This is achieved by means of
the relation $ \det (\gamma'_{kl})_1 (\gamma'^{ij})_1 = (\text{Com}
\gamma')^{ji}_1 $, valid for any matrix $ (\gamma'_{ij})_1 $ between its
determinant $ \det (\gamma'_{kl})_1$, its comatrix $ (\text{Com}
\gamma')^{ij}_1 $ and its inverse. For 3-dimensional matrices, the
determinant may be written in an Euclidean covariant form as
\begin{equation} \label{eq:det}
\det (\gamma'_{ij})_1 = \frac{1}{6} \varepsilon^{ijk} \varepsilon^{lmn}
(g'_{il})_1 (g'_{jm})_1 (g'_{kn})_1 \, .
\end{equation}
Similarly, we have for the comatrix
\begin{equation} \label{eq:com}
(\mathrm{Com} \gamma')^{ij}_1 = - \frac{1}{2} \varepsilon^{ikl} \varepsilon^{jmn}
(g'_{kn})_1 (g'_{lm})_1 \, .
\end{equation}
The inverse spatial metrics $ \gamma'^{ij}_1 $ is then given by the
ratio of the RHS of Eqs.~\eqref{eq:com} and~\eqref{eq:det}, where the
primed metric relates to $ (g_{\mu\nu})_1 $ after
Eq.~\eqref{eq:primed_metric}.

We finally look at the determination of the covariant spin components $ S'^1_i
$. As $ S^1_\mu $ is a Lorentzian vector, they are at once seen to be equal to
\begin{equation}
S'^1_i(t') = \Lambda_i^{\,\,\mu}(\mathbf{V}) S^1_{\mu}(t_1) \, ,
\end{equation}
and, by virtue of the supplementary condition~\eqref{Su}, $ S^1_0 = -
S^1_{i} v^i_1/c $.

At this stage, we have expressed $ {S'}_1^i $ in terms of quantities
evaluated at time $ t_1 $, which has led us to a relation of the form
$\mathbf{S}_1 = \boldsymbol{\mathcal{S}}_1(t_1)$. It remains to rewrite
$ \boldsymbol{\mathcal{S}}_1(t_1) $ as a function of $ t $. For the
present purpose, we restrict ourselves to a perturbative approach, and
resort to the convenient formula
\begin{equation}
f(t_1) = f(t) + \sum_{n = 1}^{+ \infty} 
\frac{(-1)^n}{c^{2n} n!} \partial_t^{n - 1} \Big[ \frac{df}{dt} 
(Vr_1)^n \Big] \, ,
\end{equation}
generalizing in a straightforward way Eq.~(3.16) of Ref.~\cite{BF01a} to
any smooth function $ f $ (see also the Appendix A of~\cite{BF01a}). In
the end, this yields the following identity for the spin ``vector'' $
\mathbf{S}'_1 = (S'^i_1)$ defined in the frame $ (\mathcal{F}') $:
\begin{equation}
\mathbf{S}'_1 = \boldsymbol{\mathcal{S}}_1(t) + \sum_{n = 1}^{+ \infty} 
\frac{(-1)^n}{c^{2n} n!} \partial_t^{n - 1} 
\bigg[ \frac{d\boldsymbol{\mathcal{S}}_1}{dt} (Vr_1)^n \bigg] \, ,
\end{equation}
where
\begin{subequations}
\begin{align} \label{eq:primed_spin}
& \mathcal{S}^i_1 = \frac{3 \delta_{ij} \Big((\tilde{g}_{kk})_1^2 -
  (\tilde{g}_{kl})_1^2 \Big) + 6 (\tilde{g}_{ik})_1 (\tilde{g}_{kj})_1  - 
6 (\tilde{g}_{kk})_1 (\tilde{g}_{ij})_1}{(\tilde{g}_{pp})_1^3 - 
3 (\tilde{g}_{pq})_1^2 (\tilde{g}_{rr})_1 +
2 (\tilde{g}_{pq})_1 (\tilde{g}_{qr})_1 (\tilde{g}_{rp})_1} \times \nonumber \\ & 
\qquad \qquad \qquad \qquad \qquad \qquad \times 
\bigg[ S_1^m (g_{jm})_1 + \gamma \frac{V^j S_1^m}{c^2} (g_{mn})_1 \Big(
-v_1^n + \frac{\gamma}{\gamma + 1} V^n \Big) \bigg] \, ,\\
& (\tilde{g}_{ij})_1 = (g_{ij})_1 + 2 \frac{\gamma V^{(i}}{c} (g_{j)0})_1 +
\gamma^2 \frac{V^i V^j}{c^2} (g_{00})_1 + \frac{2 \gamma^2}{\gamma + 1}
\frac{V^k V^{(i}}{c^2} (g_{j)k})_1 + \frac{2 \gamma^3}{\gamma + 1} 
\frac{V^i V^j V^k}{c^3} (g_{0k})_1 \nonumber \\ & \qquad \qquad \qquad \qquad
\qquad \qquad \qquad \qquad \qquad \qquad \qquad    
+ \frac{\gamma^4}{(\gamma + 1)^2} \frac{V^i V^j V^k V^l}{c^4} (g_{kl})_1 \, .
\end{align}
\end{subequations}
These expressions are valid at any order in the boost velocity
$\mathbf{V}$. After specializing the above equation truncated at the PN
level to the metric~\eqref{eq:metric2.5PN}, we arrive at
\begin{equation}
\mathbf{S}'_1 = \mathbf{S}_1 + \frac{\mathbf{V}}{c^2} \Big(-(v_1S_1) +
\frac{1}{2} (VS_1)\Big) + \mathcal{O}\Big(\frac{1}{c^4} \Big) \, .
\end{equation}
Note that all powers of $ \mathbf{V} $ consistent with the 1PN
approximation beyond the leading spin-orbit term have been included. In
principle, Eq.~\eqref{eq:primed_spin} holds only for test particles.
Nonetheless, it turns out not to depend on any regularized field. It is
thus legitimate to extend it to the conditions of the present problem.

With the previous transformation laws in hand, we are in position to check the
Lorentz invariance as explained before. After a lengthy calculation, we arrive
at the expected identity 
$ \mathbf{A}(\mathbf{y}_B, \mathbf{v}_B, \mathbf{S}_B, \delta_{ij}, 
\varepsilon_{ijk}) - \mathbf{A}'(\mathbf{y}_B, \mathbf{v}_B, \mathbf{S}_B,
\delta_{ij}, \varepsilon_{ijk}) = 0 $.

\section{Test-mass limit of the equations of motion}\label{secB} 

In the limit where one of the objects, say the number 1, is nearly at
rest while its companion has a very small mass for a finite ratio
$\mathbf{S}_2/m_2$ , we must recover the dynamics of a spinning test
particle in the background of a Kerr black hole of mass $ m_1 $ and spin
$ S_1 = m_1 a_1$ (in this Appendix we pose $G=c=1$). To allow direct
comparison with the PN equations of motion for $ m_2 \to 0 $ at
$\mathbf{S}_2/m_2 = \text{const} $, we shall work with the Kerr metric
in harmonic coordinates. The link between the Boyer-Lindquist grid
(indicated by the label BL henceforth) and some spatial harmonic
coordinates can be obtained from Eqs. (41) and (43) of Ref.~\cite{CS97}:
\begin{subequations}
\label{eq:x_harmonic}
\begin{align}
& x^1 + i x^2 = \Big(r_\text{BL} - m_1 + i a_1 \Big) \sin \theta_\text{BL}
\exp i \bigg[ \phi + \frac{a_1}{r_+ - r_-} \ln \Big| \frac{r_\text{BL} -
r_+}{r_\text{BL} - r_-} \Big| \bigg] \, ,\\ & x^3 = \Big( r_\text{BL} -
m_1 \Big) \cos \theta_\text{BL} \, ,
\end{align}
\end{subequations}
with $r_{\pm} = m_1 \pm \sqrt{m_1^2 - a_1^2} $ and $ i^2 = -1 $. Since
$\nabla^\mu \nabla_\mu t_\text{BL} = 0 $, we may also choose $ t =
t_\text{BL} $. The exact expression of the metric in the new grid is
rather complicated, but we shall not need it beyond the linear order in
the spin. Neglecting the quadratic terms $ \mathcal{O}(S_1^2) $, the
line element reduces to
\begin{align} \label{eq:Kerr_metric}
  ds^2 &= - \frac{r - m_1}{r + m_1} dt^2 - \frac{4 m_1 a_1}{r + m_1}
    \sin^2 \theta dt d\phi + \frac{r + m_1}{r - m_1} dr^2 \nonumber \\ &
    - 2 \frac{m_1^2 a_1}{r^2} \,\frac{r + m_1}{r - m_1} \sin^2 \theta dr
    d\phi + \left(r + m_1\right)^2 (d\theta^2 + \sin^2 \theta d\phi^2) +
    \mathcal{O}(S_1^2) \, ,
\end{align}
which coincides with the one deriving from the
metric~\eqref{eq:metric2.5PN} at the dominant order, hence the harmonic
coordinates defined by Eqs.~\eqref{eq:x_harmonic} and $ t = t_\text{BL}
$ are the same as those of the PN formalism.

At this level, we may derive the equations of motion of a test particle
with spin per unit mass $ S_2/m_2 $ orbiting in the gravitational
field~\eqref{eq:Kerr_metric}. For simplicity, we assume the trajectory
to be circular and lie in the equatorial plane $ \theta = \pi/2$; the
vector $ \boldsymbol{\partial}_z $ points to the direction of the spin
black hole, so that $ S_1 = S_1^z = m_1 a_1$; the spherical coordinate
basis is denoted by $(\boldsymbol{\partial}_r,
\boldsymbol{\partial}_\theta, \boldsymbol{\partial}_\phi) $. The
circularity conditions state in particular that $ r $ remains constant
in time. The spatial components of the four-velocity are then
\begin{subequations}
\begin{align}
& u^r = \frac{dr}{d\tau} = 0 \, ,\\
& u^\theta = \frac{d\theta}{d\tau} = 0 \, ,\\
& u^\phi = \frac{d\phi}{d\tau} = u^0 \frac{d\phi}{dt} \, .
\end{align}
\end{subequations}
After taking these relations into account, the explicit form of the
evolution equations~\eqref{dPAdtnew} becomes
\begin{subequations}
\begin{align} \label{eq:EOM_Kerr_1} 
& \frac{d}{dt} \bigg[u^0 \Big( g_{00} + g_{0\phi} \frac{d\phi}{dt} \Big)
\bigg] = 0 \, , \\ & \frac{d}{dt} \bigg[ u^0 g_{r\phi} \frac{d\phi}{dt}
\bigg] = (u^0)^2 \bigg[ \left(r + m_1\right)
\Big(\frac{d\phi}{dt}\Big)^2 \nonumber \\ \label{eq:EOM_Kerr_2} & \qquad
\qquad \qquad + \frac{m_1}{(r + m_1)^2} \Big(-1 + \frac{d\phi}{dt} \Big(
2 a_1 - \frac{3}{r + m_1} \frac{S^2_\theta}{m_2} \Big) \Big) \bigg] \, ,
\\
\label{eq:EOM_Kerr_3} & 0 = \frac{d\phi}{dt} \frac{m_1}{r} 
\frac{1 - m_1/r}{(1 + m_1/r)^2} \frac{S^2_r}{m_2} (u^0)^2 \, , \\
\label{eq:EOM_Kerr_4} & \frac{d}{dt} \bigg[ u^0 \Big(g_{\phi 0} +
g_{\phi\phi} \frac{d\phi}{dt}\Big) \bigg] = 0 \, .
\end{align}
\end{subequations}
The harmonic gravitational field only depends on $ r $ and $ \theta $,
both of which do not change with time. It is itself independent of $ t
$. Thus, Eqs.~\eqref{eq:EOM_Kerr_1} and~\eqref{eq:EOM_Kerr_4} imply that
$u^0 $ and $ \omega = d\phi/dt $ are constant,
whereas~\eqref{eq:EOM_Kerr_3} yields $ S^2_r = 0
$;~\eqref{eq:EOM_Kerr_2} shows that $S^2_\theta = \text{const}$ and
fixes the value of $ \omega $. We draw the time variation of the spin $
\mathbf{S}_2 $ from the precession equation~\eqref{parallel} specialized
to the Kerr background~\eqref{eq:Kerr_metric}:
\begin{subequations}
\begin{align} \label{eq:precession_Kerr}
& \frac{dS^2_r}{d\tau} = u^0 \bigg[\frac{1}{r + m_1} \frac{d\phi}{dt}
S^2_\phi + \frac{m_1}{r^2 - m_1^2} S^2_0 \bigg] \, , \\ &
\frac{dS^2_\theta}{d\tau} = 0 \, ,\\ & \frac{dS^2_\phi}{d\tau} = -
\left(r - m_1\right) \frac{d\phi}{dt} S^2_r u^0 \, .
\end{align}
\end{subequations}
Noticing that $ dS^2_r/d\tau = 0 $, it is immediate to see
from~\eqref{eq:precession_Kerr} together with the condition $ S^2_0 u^0
= - S^2_\phi u^\phi $ that $ S^2_\phi = 0$. The remaining equations are
identically satisfied. As a result, the spin of the small object is
aligned (or anti-aligned) with the spin of the black hole, meaning that
\begin{equation}
\mathbf{S}_2 = S_2^\theta \boldsymbol{\partial}_\theta = - \frac{r}{(r +
  m_1)^2} S^2_\theta \boldsymbol{\partial}_z
\end{equation}
up to possible quadratic contributions. In the test particle limit, the
spin vectors are related to $ \mathbf{S} $ and $ \mathbf{\Sigma} $ as $
\mathbf{S}_1 = \mathbf{S} + \mathcal{O}(m_2)$ and $ \mathbf{S}_2/m_2 = (
\mathbf{S} + \mathbf{\Sigma})/m + \mathcal{O}(m_2)$. Insertion of these
values in Eq.~\eqref{eq:EOM_Kerr_2} leads to the solution
\begin{equation}
\omega^2 = \frac{m}{r^3} \left\{\frac{1}{(1+\gamma)^3} -
\frac{\gamma^{3/2}}{m^2 (1+\gamma)^{9/2}} \Big[5 S_z + 3 \Sigma_z + 3
\gamma (S_z + \Sigma_z) \Big] + \mathcal{O}(S^2)\right\} \, ,
\end{equation}
with $\gamma = m/r = m_1/r + \mathcal{O}(m_2)$. By expanding the latter
equality at the 2.5PN order, we recover the generalized Kepler
relation given by~\eqref{omega}--\eqref{zeta} for $ \nu \to 0 $.

\bibliography{ListeRef}

\begin{thebibliography}{58}
\expandafter\ifx\csname natexlab\endcsname\relax\def\natexlab#1{#1}\fi
\expandafter\ifx\csname bibnamefont\endcsname\relax
  \def\bibnamefont#1{#1}\fi
\expandafter\ifx\csname bibfnamefont\endcsname\relax
  \def\bibfnamefont#1{#1}\fi
\expandafter\ifx\csname citenamefont\endcsname\relax
  \def\citenamefont#1{#1}\fi
\expandafter\ifx\csname url\endcsname\relax
  \def\url#1{\texttt{#1}}\fi
\expandafter\ifx\csname urlprefix\endcsname\relax\def\urlprefix{URL }\fi
\providecommand{\bibinfo}[2]{#2}
\providecommand{\eprint}[2][]{\url{#2}}

\bibitem[{\citenamefont{Cutler et~al.}(1993)\citenamefont{Cutler, Apostolatos,
  Bildsten, Finn, Flanagan, Kennefick, Markovic, Ori, Poisson, Sussman
  et~al.}}]{3mn}
\bibinfo{author}{\bibfnamefont{C.}~\bibnamefont{Cutler}},
  \bibinfo{author}{\bibfnamefont{T.}~\bibnamefont{Apostolatos}},
  \bibinfo{author}{\bibfnamefont{L.}~\bibnamefont{Bildsten}},
  \bibinfo{author}{\bibfnamefont{L.}~\bibnamefont{Finn}},
  \bibinfo{author}{\bibfnamefont{E.}~\bibnamefont{Flanagan}},
  \bibinfo{author}{\bibfnamefont{D.}~\bibnamefont{Kennefick}},
  \bibinfo{author}{\bibfnamefont{D.}~\bibnamefont{Markovic}},
  \bibinfo{author}{\bibfnamefont{A.}~\bibnamefont{Ori}},
  \bibinfo{author}{\bibfnamefont{E.}~\bibnamefont{Poisson}},
  \bibinfo{author}{\bibfnamefont{G.}~\bibnamefont{Sussman}},
  \bibnamefont{et~al.}, \bibinfo{journal}{Phys. Rev. Lett.}
  \textbf{\bibinfo{volume}{70}}, \bibinfo{pages}{2984} (\bibinfo{year}{1993}).

\bibitem[{\citenamefont{Cutler and Flanagan}(1994)}]{CF94}
\bibinfo{author}{\bibfnamefont{C.}~\bibnamefont{Cutler}} \bibnamefont{and}
  \bibinfo{author}{\bibfnamefont{E.}~\bibnamefont{Flanagan}},
  \bibinfo{journal}{Phys. Rev. D} \textbf{\bibinfo{volume}{49}},
  \bibinfo{pages}{2658} (\bibinfo{year}{1994}).

\bibitem[{\citenamefont{Tagoshi and Nakamura}(1994)}]{TNaka94}
\bibinfo{author}{\bibfnamefont{H.}~\bibnamefont{Tagoshi}} \bibnamefont{and}
  \bibinfo{author}{\bibfnamefont{T.}~\bibnamefont{Nakamura}},
  \bibinfo{journal}{Phys. Rev. D} \textbf{\bibinfo{volume}{49}},
  \bibinfo{pages}{4016} (\bibinfo{year}{1994}).

\bibitem[{\citenamefont{Tagoshi and Sasaki}(1994)}]{TSasa94}
\bibinfo{author}{\bibfnamefont{H.}~\bibnamefont{Tagoshi}} \bibnamefont{and}
  \bibinfo{author}{\bibfnamefont{M.}~\bibnamefont{Sasaki}},
  \bibinfo{journal}{Prog. Theor. Phys.} \textbf{\bibinfo{volume}{92}},
  \bibinfo{pages}{745} (\bibinfo{year}{1994}).

\bibitem[{\citenamefont{Damour et~al.}(1998)\citenamefont{Damour, Iyer, and
  Sathyaprakash}}]{DIS98}
\bibinfo{author}{\bibfnamefont{T.}~\bibnamefont{Damour}},
  \bibinfo{author}{\bibfnamefont{B.~R.} \bibnamefont{Iyer}}, \bibnamefont{and}
  \bibinfo{author}{\bibfnamefont{B.}~\bibnamefont{Sathyaprakash}},
  \bibinfo{journal}{Phys. Rev. D} \textbf{\bibinfo{volume}{57}},
  \bibinfo{pages}{885} (\bibinfo{year}{1998}), \eprint{gr-qc/9708034}.

\bibitem[{\citenamefont{Blanchet}(2006)}]{Bliving}
\bibinfo{author}{\bibfnamefont{L.}~\bibnamefont{Blanchet}},
  \bibinfo{journal}{Living Rev. Rel.} \textbf{\bibinfo{volume}{9}},
  \bibinfo{pages}{4} (\bibinfo{year}{2006}), \eprint{gr-qc/0202016}.

\bibitem[{\citenamefont{Buonanno
  et~al.}(2003{\natexlab{a}})\citenamefont{Buonanno, Chen, and
  Vallisneri}}]{BCV03a}
\bibinfo{author}{\bibfnamefont{A.}~\bibnamefont{Buonanno}},
  \bibinfo{author}{\bibfnamefont{Y.}~\bibnamefont{Chen}}, \bibnamefont{and}
  \bibinfo{author}{\bibfnamefont{M.}~\bibnamefont{Vallisneri}},
  \bibinfo{journal}{Phys. Rev. D} \textbf{\bibinfo{volume}{67}},
  \bibinfo{pages}{024016} (\bibinfo{year}{2003}{\natexlab{a}}),
  \bibinfo{note}{erratum Phys. Rev. D {\bf 74}, 029903 (2006)},
  \eprint{gr-qc/0205122}.

\bibitem[{\citenamefont{Buonanno
  et~al.}(2003{\natexlab{b}})\citenamefont{Buonanno, Chen, and
  Vallisneri}}]{BCV03b}
\bibinfo{author}{\bibfnamefont{A.}~\bibnamefont{Buonanno}},
  \bibinfo{author}{\bibfnamefont{Y.}~\bibnamefont{Chen}}, \bibnamefont{and}
  \bibinfo{author}{\bibfnamefont{M.}~\bibnamefont{Vallisneri}},
  \bibinfo{journal}{Phys. Rev. D} \textbf{\bibinfo{volume}{67}},
  \bibinfo{pages}{104025} (\bibinfo{year}{2003}{\natexlab{b}}),
  \eprint{gr-qc/0211087}.

\bibitem[{\citenamefont{Arun et~al.}(2005)\citenamefont{Arun, Iyer,
  Sathyaprakash, and Sundararajan}}]{AISS05}
\bibinfo{author}{\bibfnamefont{K.}~\bibnamefont{Arun}},
  \bibinfo{author}{\bibfnamefont{B.}~\bibnamefont{Iyer}},
  \bibinfo{author}{\bibfnamefont{B.}~\bibnamefont{Sathyaprakash}},
  \bibnamefont{and}
  \bibinfo{author}{\bibfnamefont{P.}~\bibnamefont{Sundararajan}},
  \bibinfo{journal}{Phys. Rev. D} \textbf{\bibinfo{volume}{71}},
  \bibinfo{pages}{084008} (\bibinfo{year}{2005}), \bibinfo{note}{erratum Phys.
  Rev. D {\bf 72}, 069903 (2005)}, \eprint{gr-qc/0411146}.

\bibitem[{\citenamefont{Blanchet et~al.}(2002)\citenamefont{Blanchet, Faye,
  Iyer, and Joguet}}]{BFIJ02}
\bibinfo{author}{\bibfnamefont{L.}~\bibnamefont{Blanchet}},
  \bibinfo{author}{\bibfnamefont{G.}~\bibnamefont{Faye}},
  \bibinfo{author}{\bibfnamefont{B.~R.} \bibnamefont{Iyer}}, \bibnamefont{and}
  \bibinfo{author}{\bibfnamefont{B.}~\bibnamefont{Joguet}},
  \bibinfo{journal}{Phys. Rev. D} \textbf{\bibinfo{volume}{65}},
  \bibinfo{pages}{061501(R)} (\bibinfo{year}{2002}), \bibinfo{note}{erratum
  Phys. Rev. D {\bf 71}, 129902(E) (2005)}, \eprint{gr-qc/0105099}.

\bibitem[{\citenamefont{Arun et~al.}(2004)\citenamefont{Arun, Blanchet, Iyer,
  and Qusailah}}]{ABIQ04}
\bibinfo{author}{\bibfnamefont{K.}~\bibnamefont{Arun}},
  \bibinfo{author}{\bibfnamefont{L.}~\bibnamefont{Blanchet}},
  \bibinfo{author}{\bibfnamefont{B.~R.} \bibnamefont{Iyer}}, \bibnamefont{and}
  \bibinfo{author}{\bibfnamefont{M.~S.} \bibnamefont{Qusailah}},
  \bibinfo{journal}{Class. Quant. Grav.} \textbf{\bibinfo{volume}{21}},
  \bibinfo{pages}{3771} (\bibinfo{year}{2004}), \bibinfo{note}{erratum Class.
  Quant. Grav. {\bf 22}, 3115 (2005)}, \eprint{gr-qc/0404085}.

\bibitem[{\citenamefont{Blanchet et~al.}(2004)\citenamefont{Blanchet, Damour,
  Esposito-Far{\`e}se, and Iyer}}]{BDEI04}
\bibinfo{author}{\bibfnamefont{L.}~\bibnamefont{Blanchet}},
  \bibinfo{author}{\bibfnamefont{T.}~\bibnamefont{Damour}},
  \bibinfo{author}{\bibfnamefont{G.}~\bibnamefont{Esposito-Far{\`e}se}},
  \bibnamefont{and} \bibinfo{author}{\bibfnamefont{B.~R.} \bibnamefont{Iyer}},
  \bibinfo{journal}{Phys. Rev. Lett.} \textbf{\bibinfo{volume}{93}},
  \bibinfo{pages}{091101} (\bibinfo{year}{2004}), \eprint{gr-qc/0406012}.

\bibitem[{\citenamefont{Abramowicz and Klu\'zniak}(2001)}]{AK01}
\bibinfo{author}{\bibfnamefont{M.~A.} \bibnamefont{Abramowicz}}
  \bibnamefont{and}
  \bibinfo{author}{\bibfnamefont{W.}~\bibnamefont{Klu\'zniak}},
  \bibinfo{journal}{Astron. Astrophys.} \textbf{\bibinfo{volume}{374}},
  \bibinfo{pages}{L19} (\bibinfo{year}{2001}), \eprint{astro-ph/0105077}.

\bibitem[{\citenamefont{Strohmayer}(2001)}]{Stro01}
\bibinfo{author}{\bibfnamefont{T.~E.} \bibnamefont{Strohmayer}},
  \bibinfo{journal}{Astrophys. J.} \textbf{\bibinfo{volume}{552}},
  \bibinfo{pages}{L49} (\bibinfo{year}{2001}), \eprint{astro-ph/0104487}.

\bibitem[{\citenamefont{Gierlinski and Done}(2004)}]{GD04}
\bibinfo{author}{\bibfnamefont{M.}~\bibnamefont{Gierlinski}} \bibnamefont{and}
  \bibinfo{author}{\bibfnamefont{C.}~\bibnamefont{Done}},
  \bibinfo{journal}{Mon. Not. R. Astron. Soc.} \textbf{\bibinfo{volume}{347}},
  \bibinfo{pages}{885} (\bibinfo{year}{2004}), \eprint{astro-ph/0307333}.

\bibitem[{\citenamefont{Fabian and Miniutti}(2005)}]{FM05}
\bibinfo{author}{\bibfnamefont{A.~C.} \bibnamefont{Fabian}} \bibnamefont{and}
  \bibinfo{author}{\bibfnamefont{G.}~\bibnamefont{Miniutti}}, in
  \emph{\bibinfo{booktitle}{Kerr Spacetime: Rotating Black Holes in General
  Relativity}}, edited by \bibinfo{editor}{\bibfnamefont{D.~L.}
  \bibnamefont{Wiltshire}},
  \bibinfo{editor}{\bibfnamefont{M.}~\bibnamefont{Visser}}, \bibnamefont{and}
  \bibinfo{editor}{\bibfnamefont{S.}~\bibnamefont{Scott}}
  (\bibinfo{publisher}{Cambridge Univ. Press}, \bibinfo{year}{2005}),
  \eprint{astro-ph/0507409}.

\bibitem[{\citenamefont{Levin and Beloborodov}(2003)}]{LB03}
\bibinfo{author}{\bibfnamefont{Y.}~\bibnamefont{Levin}} \bibnamefont{and}
  \bibinfo{author}{\bibfnamefont{A.~M.} \bibnamefont{Beloborodov}},
  \bibinfo{journal}{Astrophys. J.} \textbf{\bibinfo{volume}{590}},
  \bibinfo{pages}{L33} (\bibinfo{year}{2003}), \eprint{astro-ph/0303436}.

\bibitem[{\citenamefont{Tanaka et~al.}(1995)\citenamefont{Tanaka, Nandra,
  Fabian, Inoue, Otani, Donati, Hayashida, Iwasawa, Kii, Kunieda
  et~al.}}]{TNFIODHIKKMMM95}
\bibinfo{author}{\bibfnamefont{Y.}~\bibnamefont{Tanaka}},
  \bibinfo{author}{\bibfnamefont{K.}~\bibnamefont{Nandra}},
  \bibinfo{author}{\bibfnamefont{A.~C.} \bibnamefont{Fabian}},
  \bibinfo{author}{\bibfnamefont{H.}~\bibnamefont{Inoue}},
  \bibinfo{author}{\bibfnamefont{C.}~\bibnamefont{Otani}},
  \bibinfo{author}{\bibfnamefont{T.}~\bibnamefont{Donati}},
  \bibinfo{author}{\bibfnamefont{K.}~\bibnamefont{Hayashida}},
  \bibinfo{author}{\bibfnamefont{K.}~\bibnamefont{Iwasawa}},
  \bibinfo{author}{\bibfnamefont{T.}~\bibnamefont{Kii}},
  \bibinfo{author}{\bibfnamefont{H.}~\bibnamefont{Kunieda}},
  \bibnamefont{et~al.}, \bibinfo{journal}{Nature}
  \textbf{\bibinfo{volume}{375(6533)}}, \bibinfo{pages}{659}
  (\bibinfo{year}{1995}).

\bibitem[{\citenamefont{O'Shaughnessy et~al.}(2005)\citenamefont{O'Shaughnessy,
  Kaplan, Kalogera, and Belczynski}}]{SKKB05}
\bibinfo{author}{\bibfnamefont{R.}~\bibnamefont{O'Shaughnessy}},
  \bibinfo{author}{\bibfnamefont{J.}~\bibnamefont{Kaplan}},
  \bibinfo{author}{\bibfnamefont{V.}~\bibnamefont{Kalogera}}, \bibnamefont{and}
  \bibinfo{author}{\bibfnamefont{K.}~\bibnamefont{Belczynski}},
  \bibinfo{journal}{Astrophys. J.} \textbf{\bibinfo{volume}{632}},
  \bibinfo{pages}{1035} (\bibinfo{year}{2005}), \eprint{astro-ph/0503219}.

\bibitem[{\citenamefont{Papapetrou}(1951{\natexlab{a}})}]{Papa51}
\bibinfo{author}{\bibfnamefont{A.}~\bibnamefont{Papapetrou}},
  \bibinfo{journal}{Proc. Phys. Soc. A} \textbf{\bibinfo{volume}{64}},
  \bibinfo{pages}{57} (\bibinfo{year}{1951}{\natexlab{a}}).

\bibitem[{\citenamefont{Papapetrou}(1951{\natexlab{b}})}]{Papa51spin}
\bibinfo{author}{\bibfnamefont{A.}~\bibnamefont{Papapetrou}},
  \bibinfo{journal}{Proc. R. Soc. London A} \textbf{\bibinfo{volume}{209}},
  \bibinfo{pages}{248} (\bibinfo{year}{1951}{\natexlab{b}}).

\bibitem[{\citenamefont{Corinaldesi and Papapetrou}(1951)}]{CPapa51spin}
\bibinfo{author}{\bibfnamefont{E.}~\bibnamefont{Corinaldesi}} \bibnamefont{and}
  \bibinfo{author}{\bibfnamefont{A.}~\bibnamefont{Papapetrou}},
  \bibinfo{journal}{Proc. R. Soc. London A} \textbf{\bibinfo{volume}{209}},
  \bibinfo{pages}{259} (\bibinfo{year}{1951}).

\bibitem[{\citenamefont{Barker and O'Connell}(1975)}]{BOC75}
\bibinfo{author}{\bibfnamefont{B.}~\bibnamefont{Barker}} \bibnamefont{and}
  \bibinfo{author}{\bibfnamefont{R.}~\bibnamefont{O'Connell}},
  \bibinfo{journal}{Phys. Rev. D} \textbf{\bibinfo{volume}{12}},
  \bibinfo{pages}{329} (\bibinfo{year}{1975}).

\bibitem[{\citenamefont{Barker and O'Connell}(1979)}]{BOC79}
\bibinfo{author}{\bibfnamefont{B.}~\bibnamefont{Barker}} \bibnamefont{and}
  \bibinfo{author}{\bibfnamefont{R.}~\bibnamefont{O'Connell}},
  \bibinfo{journal}{Gen. Relativ. Gravit.} \textbf{\bibinfo{volume}{11}},
  \bibinfo{pages}{149} (\bibinfo{year}{1979}).

\bibitem[{\citenamefont{Goldberger and Rothstein}(2006)}]{GR06}
\bibinfo{author}{\bibfnamefont{W.~D.} \bibnamefont{Goldberger}}
  \bibnamefont{and} \bibinfo{author}{\bibfnamefont{I.~Z.}
  \bibnamefont{Rothstein}}, \bibinfo{journal}{Phys. Rev. D}
  \textbf{\bibinfo{volume}{73}}, \bibinfo{pages}{104029}
  (\bibinfo{year}{2006}), \eprint{hep-th/0409156}.

\bibitem[{\citenamefont{Porto}(2006)}]{Porto06}
\bibinfo{author}{\bibfnamefont{R.}~\bibnamefont{Porto}},
  \bibinfo{journal}{Phys. Rev. D} \textbf{\bibinfo{volume}{73}},
  \bibinfo{pages}{104031} (\bibinfo{year}{2006}), \eprint{gr-qc/0511061}.

\bibitem[{\citenamefont{Porto and Rothstein}(2006)}]{PR06}
\bibinfo{author}{\bibfnamefont{R.}~\bibnamefont{Porto}} \bibnamefont{and}
  \bibinfo{author}{\bibfnamefont{I.~Z.} \bibnamefont{Rothstein}},
  \bibinfo{journal}{Phys. Rev. Lett.} \textbf{\bibinfo{volume}{97}},
  \bibinfo{pages}{021101} (\bibinfo{year}{2006}), \eprint{gr-qc/0604099}.

\bibitem[{\citenamefont{Kidder et~al.}(1993)\citenamefont{Kidder, Will, and
  Wiseman}}]{KWWi93}
\bibinfo{author}{\bibfnamefont{L.}~\bibnamefont{Kidder}},
  \bibinfo{author}{\bibfnamefont{C.}~\bibnamefont{Will}}, \bibnamefont{and}
  \bibinfo{author}{\bibfnamefont{A.}~\bibnamefont{Wiseman}},
  \bibinfo{journal}{Phys. Rev. D} \textbf{\bibinfo{volume}{47}},
  \bibinfo{pages}{R4183} (\bibinfo{year}{1993}).

\bibitem[{\citenamefont{Kidder}(1995)}]{K95}
\bibinfo{author}{\bibfnamefont{L.}~\bibnamefont{Kidder}},
  \bibinfo{journal}{Phys. Rev. D} \textbf{\bibinfo{volume}{52}},
  \bibinfo{pages}{821} (\bibinfo{year}{1995}), \eprint{gr-qc/9506022}.

\bibitem[{\citenamefont{Gergely}(1999)}]{Ger99}
\bibinfo{author}{\bibfnamefont{L.}~\bibnamefont{Gergely}},
  \bibinfo{journal}{Phys. Rev. D} \textbf{\bibinfo{volume}{61}},
  \bibinfo{pages}{024035} (\bibinfo{year}{1999}), \eprint{gr-qc/9911082}.

\bibitem[{\citenamefont{Mik\'oczi et~al.}(2005)\citenamefont{Mik\'oczi,
  Vas\'uth, and Gergely}}]{MVGer05}
\bibinfo{author}{\bibfnamefont{B.}~\bibnamefont{Mik\'oczi}},
  \bibinfo{author}{\bibfnamefont{M.}~\bibnamefont{Vas\'uth}}, \bibnamefont{and}
  \bibinfo{author}{\bibfnamefont{L.}~\bibnamefont{Gergely}},
  \bibinfo{journal}{Phys. Rev. D} \textbf{\bibinfo{volume}{71}},
  \bibinfo{pages}{124043} (\bibinfo{year}{2005}), \eprint{astro-ph/0504538}.

\bibitem[{\citenamefont{Apostolatos et~al.}(1994)\citenamefont{Apostolatos,
  Cutler, Sussman, and Thorne}}]{ACST94}
\bibinfo{author}{\bibfnamefont{T.}~\bibnamefont{Apostolatos}},
  \bibinfo{author}{\bibfnamefont{C.}~\bibnamefont{Cutler}},
  \bibinfo{author}{\bibfnamefont{G.}~\bibnamefont{Sussman}}, \bibnamefont{and}
  \bibinfo{author}{\bibfnamefont{K.}~\bibnamefont{Thorne}},
  \bibinfo{journal}{Phys. Rev. D} \textbf{\bibinfo{volume}{49}},
  \bibinfo{pages}{6274} (\bibinfo{year}{1994}).

\bibitem[{\citenamefont{Pan et~al.}(2004)\citenamefont{Pan, Buonanno, Chen, and
  Vallisneri}}]{PBCV04}
\bibinfo{author}{\bibfnamefont{Y.}~\bibnamefont{Pan}},
  \bibinfo{author}{\bibfnamefont{A.}~\bibnamefont{Buonanno}},
  \bibinfo{author}{\bibfnamefont{Y.}~\bibnamefont{Chen}}, \bibnamefont{and}
  \bibinfo{author}{\bibfnamefont{M.}~\bibnamefont{Vallisneri}},
  \bibinfo{journal}{Phys. Rev. D} \textbf{\bibinfo{volume}{69}},
  \bibinfo{pages}{104017} (\bibinfo{year}{2004}), \eprint{gr-qc/0310034}.

\bibitem[{\citenamefont{Buonanno et~al.}(2004)\citenamefont{Buonanno, Chen,
  Pan, and Vallisneri}}]{BCPV04}
\bibinfo{author}{\bibfnamefont{A.}~\bibnamefont{Buonanno}},
  \bibinfo{author}{\bibfnamefont{Y.}~\bibnamefont{Chen}},
  \bibinfo{author}{\bibfnamefont{Y.}~\bibnamefont{Pan}}, \bibnamefont{and}
  \bibinfo{author}{\bibfnamefont{M.}~\bibnamefont{Vallisneri}},
  \bibinfo{journal}{Phys. Rev. D} \textbf{\bibinfo{volume}{70}},
  \bibinfo{pages}{104003} (\bibinfo{year}{2004}), \bibinfo{note}{erratum Phys.
  Rev. D {\bf 74}, 029902 (2006)}, \eprint{gr-qc/0405090}.

\bibitem[{\citenamefont{Buonanno et~al.}(2005)\citenamefont{Buonanno, Chen,
  Pan, Tagoshi, and Vallisneri}}]{BCPTV05}
\bibinfo{author}{\bibfnamefont{A.}~\bibnamefont{Buonanno}},
  \bibinfo{author}{\bibfnamefont{Y.}~\bibnamefont{Chen}},
  \bibinfo{author}{\bibfnamefont{Y.}~\bibnamefont{Pan}},
  \bibinfo{author}{\bibfnamefont{H.}~\bibnamefont{Tagoshi}}, \bibnamefont{and}
  \bibinfo{author}{\bibfnamefont{M.}~\bibnamefont{Vallisneri}},
  \bibinfo{journal}{Phys. Rev. D} \textbf{\bibinfo{volume}{72}},
  \bibinfo{pages}{084027} (\bibinfo{year}{2005}), \eprint{gr-qc/0508064}.

\bibitem[{\citenamefont{Owen et~al.}(1998)\citenamefont{Owen, Tagoshi, and
  Ohashi}}]{OTO98}
\bibinfo{author}{\bibfnamefont{B.}~\bibnamefont{Owen}},
  \bibinfo{author}{\bibfnamefont{H.}~\bibnamefont{Tagoshi}}, \bibnamefont{and}
  \bibinfo{author}{\bibfnamefont{A.}~\bibnamefont{Ohashi}},
  \bibinfo{journal}{Phys. Rev. D} \textbf{\bibinfo{volume}{57}},
  \bibinfo{pages}{6168} (\bibinfo{year}{1998}), \eprint{gr-qc/9710134}.

\bibitem[{\citenamefont{Tagoshi et~al.}(2001)\citenamefont{Tagoshi, Ohashi, and
  Owen}}]{TOO01}
\bibinfo{author}{\bibfnamefont{H.}~\bibnamefont{Tagoshi}},
  \bibinfo{author}{\bibfnamefont{A.}~\bibnamefont{Ohashi}}, \bibnamefont{and}
  \bibinfo{author}{\bibfnamefont{B.}~\bibnamefont{Owen}},
  \bibinfo{journal}{Phys. Rev. D} \textbf{\bibinfo{volume}{63}},
  \bibinfo{pages}{044006} (\bibinfo{year}{2001}), \eprint{gr-qc/0010014}.

\bibitem[{\citenamefont{Blanchet et~al.}(2006)\citenamefont{Blanchet, Buonanno,
  and Faye}}]{BBF06spin}
\bibinfo{author}{\bibfnamefont{L.}~\bibnamefont{Blanchet}},
  \bibinfo{author}{\bibfnamefont{A.}~\bibnamefont{Buonanno}}, \bibnamefont{and}
  \bibinfo{author}{\bibfnamefont{G.}~\bibnamefont{Faye}},
  \bibinfo{journal}{Phys. Rev. D} \textbf{\bibinfo{volume}{74}},
  \bibinfo{pages}{104034} (\bibinfo{year}{2006}), \bibinfo{note}{erratum to be
  published in Phys. Rev. D}, \eprint{gr-qc/0605140}.

\bibitem[{\citenamefont{Tulczyjew}(1957)}]{T57}
\bibinfo{author}{\bibfnamefont{W.}~\bibnamefont{Tulczyjew}},
  \bibinfo{journal}{Bull. Acad. Polon. Sci.} \textbf{\bibinfo{volume}{III, 5}},
  \bibinfo{pages}{279} (\bibinfo{year}{1957}).

\bibitem[{\citenamefont{Tulczyjew}(1959)}]{T59}
\bibinfo{author}{\bibfnamefont{W.}~\bibnamefont{Tulczyjew}},
  \bibinfo{journal}{Acta Phys. Polon.} \textbf{\bibinfo{volume}{18}},
  \bibinfo{pages}{37} (\bibinfo{year}{1959}).

\bibitem[{\citenamefont{Trautman}(2002)}]{Traut58}
\bibinfo{author}{\bibfnamefont{A.}~\bibnamefont{Trautman}},
  \bibinfo{journal}{Gen. Relat. Grav.} \textbf{\bibinfo{volume}{34}},
  \bibinfo{pages}{721} (\bibinfo{year}{2002}), \bibinfo{note}{reprinted from
  lectures delivered in 1958}.

\bibitem[{\citenamefont{Dixon}(1979)}]{Dixon}
\bibinfo{author}{\bibfnamefont{W.}~\bibnamefont{Dixon}}, in
  \emph{\bibinfo{booktitle}{Isolated systems in general relativity}}, edited by
  \bibinfo{editor}{\bibfnamefont{J.}~\bibnamefont{Ehlers}}
  (\bibinfo{publisher}{North Holland}, \bibinfo{address}{Amsterdam},
  \bibinfo{year}{1979}), p. \bibinfo{pages}{156}.

\bibitem[{\citenamefont{Bailey and Israel}(1980)}]{BI80}
\bibinfo{author}{\bibfnamefont{I.}~\bibnamefont{Bailey}} \bibnamefont{and}
  \bibinfo{author}{\bibfnamefont{W.}~\bibnamefont{Israel}},
  \bibinfo{journal}{Ann. Phys.} \textbf{\bibinfo{volume}{130}},
  \bibinfo{pages}{188} (\bibinfo{year}{1980}).

\bibitem[{\citenamefont{Damour}(1982)}]{D82}
\bibinfo{author}{\bibfnamefont{T.}~\bibnamefont{Damour}}, \bibinfo{journal}{C.
  R. Acad. Sc. Paris} \textbf{\bibinfo{volume}{294}}, \bibinfo{pages}{1355}
  (\bibinfo{year}{1982}).

\bibitem[{\citenamefont{Mino et~al.}(1996)\citenamefont{Mino, Shibata, and
  Tanaka}}]{MST96}
\bibinfo{author}{\bibfnamefont{Y.}~\bibnamefont{Mino}},
  \bibinfo{author}{\bibfnamefont{M.}~\bibnamefont{Shibata}}, \bibnamefont{and}
  \bibinfo{author}{\bibfnamefont{T.}~\bibnamefont{Tanaka}},
  \bibinfo{journal}{Phys. Rev. D} \textbf{\bibinfo{volume}{53}},
  \bibinfo{pages}{622} (\bibinfo{year}{1996}).

\bibitem[{\citenamefont{Tanaka et~al.}(1996)\citenamefont{Tanaka, Mino, Sasaki,
  and Shibata}}]{TMSS96}
\bibinfo{author}{\bibfnamefont{T.}~\bibnamefont{Tanaka}},
  \bibinfo{author}{\bibfnamefont{Y.}~\bibnamefont{Mino}},
  \bibinfo{author}{\bibfnamefont{M.}~\bibnamefont{Sasaki}}, \bibnamefont{and}
  \bibinfo{author}{\bibnamefont{Shibata}}, \bibinfo{journal}{Phys. Rev. D}
  \textbf{\bibinfo{volume}{54}}, \bibinfo{pages}{3762} (\bibinfo{year}{1996}),
  \eprint{gr-qc/9602038}.

\bibitem[{\citenamefont{Blanchet et~al.}(1998)\citenamefont{Blanchet, Faye, and
  Ponsot}}]{BFP98}
\bibinfo{author}{\bibfnamefont{L.}~\bibnamefont{Blanchet}},
  \bibinfo{author}{\bibfnamefont{G.}~\bibnamefont{Faye}}, \bibnamefont{and}
  \bibinfo{author}{\bibfnamefont{B.}~\bibnamefont{Ponsot}},
  \bibinfo{journal}{Phys. Rev. D} \textbf{\bibinfo{volume}{58}},
  \bibinfo{pages}{124002} (\bibinfo{year}{1998}), \eprint{gr-qc/9804079}.

\bibitem[{\citenamefont{Blanchet and Faye}(2000{\natexlab{a}})}]{BFreg}
\bibinfo{author}{\bibfnamefont{L.}~\bibnamefont{Blanchet}} \bibnamefont{and}
  \bibinfo{author}{\bibfnamefont{G.}~\bibnamefont{Faye}}, \bibinfo{journal}{J.
  Math. Phys.} \textbf{\bibinfo{volume}{41}}, \bibinfo{pages}{7675}
  (\bibinfo{year}{2000}{\natexlab{a}}), \eprint{gr-qc/0004008}.

\bibitem[{\citenamefont{Blanchet and Faye}(2001{\natexlab{a}})}]{BFeom}
\bibinfo{author}{\bibfnamefont{L.}~\bibnamefont{Blanchet}} \bibnamefont{and}
  \bibinfo{author}{\bibfnamefont{G.}~\bibnamefont{Faye}},
  \bibinfo{journal}{Phys. Rev. D} \textbf{\bibinfo{volume}{63}},
  \bibinfo{pages}{062005} (\bibinfo{year}{2001}{\natexlab{a}}),
  \eprint{gr-qc/0007051}.

\bibitem[{\citenamefont{Blanchet and Faye}(2000{\natexlab{b}})}]{BF00}
\bibinfo{author}{\bibfnamefont{L.}~\bibnamefont{Blanchet}} \bibnamefont{and}
  \bibinfo{author}{\bibfnamefont{G.}~\bibnamefont{Faye}},
  \bibinfo{journal}{Phys. Lett. A} \textbf{\bibinfo{volume}{271}},
  \bibinfo{pages}{58} (\bibinfo{year}{2000}{\natexlab{b}}),
  \eprint{gr-qc/0004009}.

\bibitem[{\citenamefont{Fock}(1959)}]{Fock}
\bibinfo{author}{\bibfnamefont{V.}~\bibnamefont{Fock}},
  \emph{\bibinfo{title}{Theory of space, time and gravitation}}
  (\bibinfo{publisher}{Pergamon}, \bibinfo{address}{London},
  \bibinfo{year}{1959}).

\bibitem[{\citenamefont{Damour}(1983)}]{D83houches}
\bibinfo{author}{\bibfnamefont{T.}~\bibnamefont{Damour}}, in
  \emph{\bibinfo{booktitle}{Gravitational Radiation}}, edited by
  \bibinfo{editor}{\bibfnamefont{N.}~\bibnamefont{Deruelle}} \bibnamefont{and}
  \bibinfo{editor}{\bibfnamefont{T.}~\bibnamefont{Piran}}
  (\bibinfo{publisher}{North-Holland Company}, \bibinfo{address}{Amsterdam},
  \bibinfo{year}{1983}), pp. \bibinfo{pages}{59--144}.

\bibitem[{\citenamefont{Will}(2005)}]{W05}
\bibinfo{author}{\bibfnamefont{C.}~\bibnamefont{Will}}, \bibinfo{journal}{Phys.
  Rev. D} \textbf{\bibinfo{volume}{71}}, \bibinfo{pages}{084027}
  (\bibinfo{year}{2005}), \eprint{gr-qc/0502039}.

\bibitem[{\citenamefont{de~Andrade et~al.}(2001)\citenamefont{de~Andrade,
  Blanchet, and Faye}}]{ABF01}
\bibinfo{author}{\bibfnamefont{V.}~\bibnamefont{de~Andrade}},
  \bibinfo{author}{\bibfnamefont{L.}~\bibnamefont{Blanchet}}, \bibnamefont{and}
  \bibinfo{author}{\bibfnamefont{G.}~\bibnamefont{Faye}},
  \bibinfo{journal}{Class. Quant. Grav.} \textbf{\bibinfo{volume}{18}},
  \bibinfo{pages}{753} (\bibinfo{year}{2001}), \eprint{gr-qc/0011063}.

\bibitem[{\citenamefont{Blanchet and Iyer}(2003)}]{BI03CM}
\bibinfo{author}{\bibfnamefont{L.}~\bibnamefont{Blanchet}} \bibnamefont{and}
  \bibinfo{author}{\bibfnamefont{B.~R.} \bibnamefont{Iyer}},
  \bibinfo{journal}{Class. Quant. Grav.} \textbf{\bibinfo{volume}{20}},
  \bibinfo{pages}{755} (\bibinfo{year}{2003}), \eprint{gr-qc/0209089}.

\bibitem[{\citenamefont{Damour and Deruelle}(1981)}]{DD81b}
\bibinfo{author}{\bibfnamefont{T.}~\bibnamefont{Damour}} \bibnamefont{and}
  \bibinfo{author}{\bibfnamefont{N.}~\bibnamefont{Deruelle}},
  \bibinfo{journal}{C. R. Acad. Sc. Paris} \textbf{\bibinfo{volume}{293}},
  \bibinfo{pages}{537} (\bibinfo{year}{1981}).

\bibitem[{\citenamefont{Blanchet and Faye}(2001{\natexlab{b}})}]{BF01a}
\bibinfo{author}{\bibfnamefont{L.}~\bibnamefont{Blanchet}} \bibnamefont{and}
  \bibinfo{author}{\bibfnamefont{G.}~\bibnamefont{Faye}}, \bibinfo{journal}{J.
  Math. Phys.} \textbf{\bibinfo{volume}{42}}, \bibinfo{pages}{4391}
  (\bibinfo{year}{2001}{\natexlab{b}}), \eprint{gr-qc/0006100}.

\bibitem[{\citenamefont{Cook and Scheel}(1997)}]{CS97}
\bibinfo{author}{\bibfnamefont{G.~B.} \bibnamefont{Cook}} \bibnamefont{and}
  \bibinfo{author}{\bibfnamefont{M.~A.} \bibnamefont{Scheel}},
  \bibinfo{journal}{Phys. Rev. D} \textbf{\bibinfo{volume}{56}},
  \bibinfo{pages}{4775} (\bibinfo{year}{1997}).

\end{thebibliography}

\end{document}